# Microscale bending plasticity and fracture behavior of amorphous aluminum oxide films

**Nidhin George Mathews[1,2,a,†], Erkka J. Frankberg[1,a,†], Vivek Devulapalli[3], Chandan Kumar[1], Barbara Putz[3,4], Aloshious Lambai[1,5], Sergei Khakalo[6], Mattia Cabrioli[7], Bjarke Holl Christensen[8], Janne-Petteri Niemelä[3], Arnold Milenko Müller[9], Fabio Di Fonzo[7,10], Ivo Utke[3], Erkki Levänen[1], Gaurav Mohanty[1]**

[1] *Materials Science and Environmental Engineering, Faculty of Engineering and Natural Sciences, Tampere University, 33720 Tampere, Finland*

[2] *Advanced Materials for Nuclear Energy, VTT Technical Research Centre of Finland, 02150 Espoo, Finland*

[3] *Laboratory for Mechanics of Materials and Nanostructures, Empa-Swiss Federal Laboratories for Materials Science and Technology, 3602 Thun, Switzerland*

[4] *Department of Materials Science, University of Leoben, 8700 Leoben, Austria*

[5] *Materials for Emerging Technologies, VTT Technical Research Centre of Finland, 02150 Espoo, Finland*

[6] *Department of Civil Engineering, Aalto University, 02150 Espoo, Finland*

[7] *Center for Nano Science and Technology CNST @ Polimi, Istituto Italiano di Tecnologia, 20134 Milano, Italy*

[8] *Danish Technological Institute, 8000 Aarhus, Denmark*

[9] *Laboratory of Ion Beam Physics, ETH Zürich, 8093 Zürich, Switzerland.*

[10] *X-nano s.r.l., 20124 Milano, Italy*

[a] *Corresponding author: nidhin.mathews@tuni.fi, erkka.frankberg@tuni.fi*

[†] *These authors contributed equally to the paper writing.*

**Abstract**

Recent work has demonstrated microscale compressive plasticity in pulse laser deposited (PLD) amorphous alumina (a-$Al_2O_3$). This work explores microscale bending plasticity and fracture behavior of a-$Al_2O_3$ films deposited using three different methods – PLD, atomic layer deposition (ALD) and sputter deposition (SD). The three deposition routes produced amorphous films with similar stoichiometric compositions. We demonstrate, for the first time, bending plasticity in PLD and ALD a-$Al_2O_3$ films at microscale using *in situ* microcantilever bending experiments at room temperature. All tested PLD a-$Al_2O_3$ microcantilevers showed substantial ductile behavior in bending by accommodating total strains > 10% without fracture. Half of the tested ALD a-$Al_2O_3$ cantilevers exhibited elastic brittle fracture while the other half showed bending plasticity, indicating that the observed deformation behavior is strongly influenced by the presence and distribution of defects within the tested volume. All SD a-$Al_2O_3$ microcantilevers showed elastic brittle failure attributed to their columnar growth microstructure. The microscale bending response was found to be highly dependent on the film deposition method highlighting the role of defects in suppressing plasticity mechanisms. Notched microcantilever bending tests on all three films showed brittle failure with similar fracture toughness value of 3.1 ± 0.2 MPa.m$^{0.5}$, effectively ruling out any localized crack tip plasticity. These findings underscore the importance of minimizing defects during fabrication in order to develop damage tolerant amorphous oxides. Nonetheless, the observation of bending plasticity in both PLD and ALD microcantilevers, which include a tensile stress component as well, suggests that the plastic deformation mechanisms in amorphous alumina are more general and are not exclusively governed by the deposition method. This opens promising possibilities for their use in engineering applications.

# 1 Introduction

Generally, oxide materials are known to be brittle due to lack of active plasticity mechanisms at room temperature, and this inherent brittleness limits their usage in many modern technologies. However, recent reports suggest that amorphous oxide materials can show plasticity at small length scales [1–4]. Specifically, amorphous aluminum oxide (a-$Al_2O_3$), an oxide glass material, has been shown to exhibit room temperature plasticity at nano- and micrometer length scales, and there is a high degree of interest in better understanding the origin of its exceptional mechanical behavior and its limits, in terms of loading conditions and length scales. Amorphous $Al_2O_3$ shows promising potential for applications in modern electronics [5,6], hydrogen barrier coatings [7,8], thermal oxidation barrier coatings [9], energy storage technologies [10,11] and flexible electronic systems [12]. Previous study by Frankberg et al. [3] reported that nanoscale thin films of a-$Al_2O_3$, made by pulsed laser deposition (PLD), exhibit significant unconfined plasticity at room temperature under all principal loading modes. They observed ~ 5-8% plastic strain in tension using *in situ* TEM experiments and proposed that the observed viscous creep plasticity could, in principle, be extended to micrometer length scales and beyond, if the material can be fabricated in fully dense and flaw-free form. More recently, Frankberg et al. reported *in situ* SEM micropillar compression of the same material - PLD a-$Al_2O_3$ films - showing substantial microscale plasticity. Micropillars of 2 μm diameter and 6 μm height could be compressed up to ~ 50% total strain without fracture, over 6 orders of strain rates, from quasistatic $10^{-3}$ $s^{-1}$ up to impact-type loading at $10^{3}$ $s^{-1}$ [4]. The exceptional compression plasticity of PLD a-$Al_2O_3$ can be gauged from the fact that literature reports microcompression of other ceramic systems to be typically < 10% plastic strain-to-failure. Both these studies ruled out the influence of electron beam in enhancing the plasticity of a-$Al_2O_3$ through systematic beam ON and beam OFF experiments, which did not show any significant difference. Although tensile hoop stresses are present in the micropillars during compression [2,4], the lack of catastrophic fracture did not fully confirm the role that tensile plasticity plays towards overall deformation. Questions remain on whether a-$Al_2O_3$ (and amorphous ceramics in general) shows tensile or bending plasticity at micrometer length scales and if its presence can enhance its fracture toughness values through local crack tip plasticity mechanisms. Another open question is whether a-$Al_2O_3$ films deposited using methods other than PLD, as reported earlier [3,4], show similar beneficial mechanical behavior in terms of plasticity. If this were true, it will significantly expand the scope of its applications and provide significant insights into plasticity mechanisms in ceramics as a function of microstructure and chemical composition.

Several recent studies have demonstrated small-scale room temperature plasticity in ceramic materials in compression [1,4,13] including amorphous oxide materials [14,15]. Amorphous ceramics with embedded nanoscale amorphous/crystalline particles showed enhanced plastic deformation [15–18]. However, compression typically suppresses and even closes the existing defects (pores and cracks), thereby, allowing to surpass the yield strength of the material. Therefore, tensile plasticity is a better indicator of ductility and plastic deformation. So far, tensile plasticity has been demonstrated only at the nanometer length scales (nanometric thin films, wire and ribbons) in materials such as silicon nitride, boron nitride, aluminum oxide, silicon oxide [3,19–21]. Despite the low fracture toughness typically measured for amorphous oxide materials [3,22,23], plastic deformation has been found to occur even at room temperature [1–4] in special conditions in samples of extreme nanoscale [19], or artificially densified samples [24] or the presence of an electron beam [25]. So far, pure tensile or flexural plasticity in amorphous oxide films has been achieved only for thin nanoscale samples and no results can be found at micrometer length scales [26,27,3]. Performing microscale tensile tests is extremely time consuming and expensive requiring the use of advanced femtosecond laser or plasma focused ion beam [28]. A more accessible microscale test to indirectly study tensile plasticity in small volumes is microcantilever bending test. Microcantilevers have been a widely used to characterise the plastic deformation and fracture behaviour of films and coatings [29]. During microcantilever bending, tensile stresses are generated in the region above the neutral axis, which consequently promotes tensile plasticity in case of ductile material and failure in case of brittle material. By incorporating a focused ion beam (FIB) notch near the fixed end of the cantilever, notched microcantilever bending (NCB) test can be performed to determine fracture toughness and investigate local crack tip plasticity mechanisms, if any. The formation of a plastic zone at the crack tip dissipates elastic energy and increases the fracture toughness value, typically resulting in 1-2 orders of magnitude difference in fracture toughness between brittle and ductile materials. There is lack of information on both bending behavior and fracture properties of a-$Al_2O_3$ films. The closest data available from microscale NCB tests is for its crystalline counterpart. Microscale NCB fracture toughness values of crystalline $Al_2O_3$ range from 1.4 - 4.3 $MPa.m^{0.5}$ [30–32], depending on the microstructural features such as grain orientation and presence/absence of grain boundaries. While the data on NCB fracture toughness on a-$Al_2O_3$ is clearly not yet available, fracture toughness of ALD a-$Al_2O_3$ films deposited on stretchable substrates has been estimated to range between 1.7 - 3.1 $MPa.m^{0.5}$ from shear-lag tensile straining experiments [3,22,23,33,34]. However, these tests were performed on films attached to the substrate which are subjected to residual stresses. They do not provide classical mode I fracture toughness values of the free-standing films, like NCB tests under plane strain conditions.

In this study, we attempt to rationalize the unique mechanical behavior of a-$Al_2O_3$ films by testing this material processed through three different deposition routes which can produce characteristic amorphous structure – atomic layer deposition, sputter deposition, and pulsed laser deposition. These three methods were used to grow a-$Al_2O_3$ films which were a couple of micrometers thick. We investigate the bending behaviour and fracture properties of these three a-$Al_2O_3$ film systems using both notched and unnotched microcantilever bending experiments that generate tensile stress state within the deformation volume. Our results demonstrate microscale ductile bending of an amorphous ceramic material at room temperature, which brings this functionality closer to be utilized in real engineering applications at this length-scale.

## 2 Materials and methodology

### 2.1 Deposition details of a-$Al_2O_3$ films

Aluminum oxide films were grown on single crystal silicon substrates using three different synthesis methods: i) atomic layer deposition (ALD), ii) sputter deposition (SD), and iii) pulsed laser deposition (PLD). A detailed explanation of each deposition methodology is provided in this section.

*Atomic layer deposition* runs were performed in an ALD reactor (Arradiance GEMStar XT, USA), with a precursor sequence of pulse-exposure-purge at 0.05-5-25 s for trimethylaluminium (TMA, Sigma Aldrich 97%) and 0.05-5-25 s for $H_2O$ [35]. The exhaust valve of the deposition chamber was closed during each precursor pulse and during the exposure period following the pulse. Depositions were performed at 120 °C on single silicon crystal substrates. Both precursor lines, the door and the chamber of the ALD reactor were heated to 120 °C, with a 4-hour pre-heating step prior to the deposition. A total film thickness of ~2 μm was deposited by performing 15,400 ALD cycles, with a growth-per-cycle (GPC) value of 0.13 nm.

All *sputter depositions* were made with an industrial-scale CemeCon CC800/9 SinOx sputtering system which was operated in dual pulsed mode with a repetition rate of 50 kHz and 50% duty cycle [36]. This method used reactive sputtering with a metallic Al target and injected oxygen. Metallic Al target was used because the deposition rate with an $Al_2O_3$ target is extremely low. Aluminum 1050 (99.5% purity) with the dimensions of $1*8.8*50\ cm^3$ was used as the sputtering target, and single crystal silicon was used as the substrate. The substrates were held on a planetary rotating substrate table to achieve homogeneous film depositions. A mixture of argon (99.999%) and oxygen (99.999%) was used as the process gas. Oxygen flow was regulated using a feedback loop. The chamber pressures were maintained at 0.5 mPa prior to deposition. The cathode power was fixed at 4 kW and the cathode voltage of 430 V was set during depositions. A 600 nm thick interlayer of TiAlN was deposited on the silicon substrate prior to the deposition of $Al_2O_3$ to achieve good adhesion of $Al_2O_3$ films on the substrate. The depositions were performed for a total film thickness of ~ 8 μm at temperature of 150 °C to avoid crystallization.

*Pulsed laser depositions* were performed in a custom-made vacuum chamber (I-PLD300, Kenosistec s.r.l, Italy) using a nanosecond UV laser source (Coherent Gmbh, Germany) with wavelength 248 nm directed towards a polycrystalline $Al_2O_3$ target (Testbourne 99.99% purity) [3]. Laser fluence of 3.5 J/cm$^2$ and repetition rate of 50 Hz was used with the target to the substrate distance kept at 50 mm. Several a-$Al_2O_3$ films, with thicknesses up to 10 µm, were deposited on single crystal silicon wafer substrates at room temperature. Oxygen gas was supplied to the chamber at a pressure of 0.15 Pa to maintain the required stoichiometry of the deposited a-$Al_2O_3$.

### 2.2 Structural and mechanical characterization of films

The structural characterization of the deposited films was performed using grazing incidence X-ray diffraction (GI-XRD, Malvern Panalytical Empyrean, U.K) to confirm the amorphous structure. X-ray scans were performed using Cu-$K_\alpha$ radiation (wavelength, $\lambda = 0.154$ nm) within the selected 2θ range, with stepsize of 0.016º at 40 kV operating voltage and 30 mA current. The X-ray incidence angle was selected such that the intensity of substrate peaks is minimized. The chemical composition and stoichiometry of all three films were verified by Rutherford backscattering spectrometry (RBS) using 2 MeV $He^{2+}$ ions and time-of-flight elastic recoil detection analysis (ToF-ERDA) using $^{127}I$ ions with 13 MeV energy using a 1.7 MV Tandetron accelerator. The ERDA spectra were analyzed in Potku software and depth resolved chemical composition was determined.

Cross-sectional transmission electron microscopy (TEM) specimens were prepared using focused ion beam (FIB) (Tescan Lyra, Czech Republic) with 30 kV $Ga^+$ ions followed by 5 kV cleaning. TEM analysis was performed using Thermo Fisher Titan Themis used at 200 kV for both bright field and high angle annular dark field (HAADF) scanning TEM (STEM) imaging.

Elastic modulus ($E$) and hardness ($H$) of the deposited films were determined by nanoindentation measurements using an Alemnis *in situ* SEM nanoindenter (Alemnis AG, Switzerland). Indentations were performed on all film samples using a diamond Berkovich indenter tip in depth-controlled mode to a maximum depth ($h_{max}$) of ~ 230 nm. The maximum depth of indentation lies within 10% of film thickness ($t$) to avoid any elastic substrate effects from affecting the measurements ($h_{max}/t$ for ALD = 10%, SD = 3%, PLD = 2%). The load-displacement response of all films from nanoindentation is shown in Fig. S1 (supplementary information section S1). Indentation modulus and hardness values were determined from nanoindentation load-displacement curves using the Oliver-Pharr method [37].

## 2.3 Microcantilever sample fabrication and experiments

Microscale samples required for the micromechanical bending and fracture experiments were fabricated on these films using focused ion beam (FIB) milling (Zeiss Crossbeam 540, Germany) equipped with a gallium source, operated at 30 kV voltage. Free standing rectangular microcantilevers were fabricated at the edge of the sample with target dimensions of 16∗3∗3 µm (corresponding to length ($L_1$ or $L$)∗ thickness ($W$) ∗width ($B$), see Fig.1a for definitions). Ion currents of 15 nA, 3 nA and 0.3 nA were used for coarse, intermediate, and fine milling of the cantilever specimens. The front and back surface of the cantilever were fine polished at an additional tilt of ± 1.5° to ensure uniform cross-section. The thickness ($W$) of the cantilever was measured to be between 2 - 3 µm, so accordingly, a length-to-thickness ratio ($L_1/W$) in the range of 4.5 – 5 was maintained during the experiments. $L_1$ is the cantilever length between the fixed end and the point of loading for unnotched cantilever, whereas $L$ is the length between the notch and the point of loading for notched cantilever. For microcantilever fracture specimens, notches were milled using 10 pA milling current near the fixed end of the cantilever at a distance ($x/W$) ≈ 1, where $x$ is the distance between the notch and the fixed end of the cantilever. A notch depth-to-thickness ratio ($a/W$) ≈ 0.35 was selected for the experiments, where $a$ is the notch depth (refer Fig. 1b to visualize the nomenclatures used). These geometrical dimensions of the microcantilever were selected so as to ensure dominant mode I loading conditions during fracture tests [38].

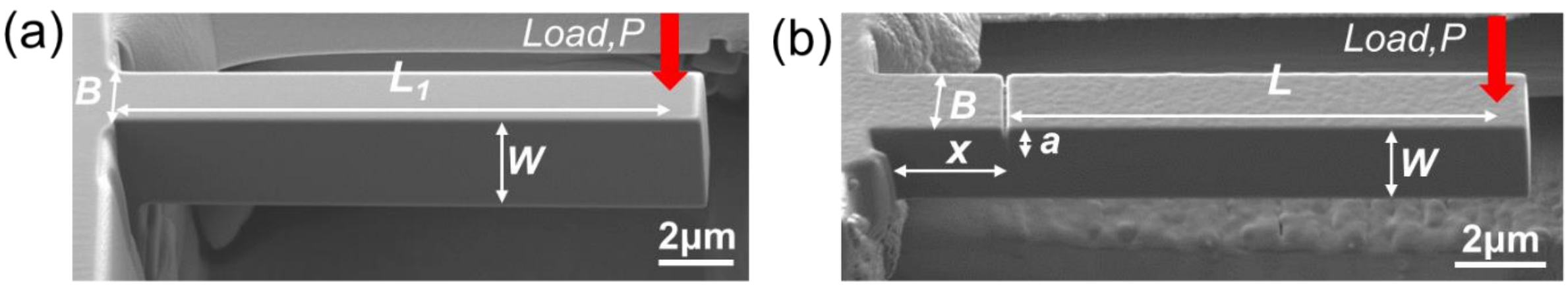

*Fig. 1: Representative image of microcantilevers used for a) bending (un-notched), and b) fracture (notched) experiments, with the nomenclatures used.*

The bending experiments were performed in displacement-controlled mode using Alemnis in-situ SEM nanoindenter (Alemnis AG, Switzerland) inside a SEM (Zeiss Leo 1450, Germany) operated at 10 kV acceleration voltage. The sample was mounted onto a SLC-0.5 (standard load cell from Alemnis) that is calibrated up to 500 mN. The displacement was applied from the tip side using a piezoactuator. For microcantilever bending and fracture experiments, a sharp diamond wedge tip (Synton MDP, Switzerland) was used with 10 μm wedge length and ~95º included angle. For both experiments, a tip displacement rate of 50 nm/s was applied. The wedge tip bends the microcantilever at a position close to its free end while maintaining the predefined length-to-thickness ratio ($L_1/W$ for bending, $L/W$ for fracture experiments). At least four microcantilevers were tested for each of the ALD, SD and PLD film systems.

For cantilever bending experiments, the applied force was converted to bending stress for normalization across samples. Bending stresses ($\sigma_b$) at the top layer of the cantilevers were estimated using the concept of simple beam bending theory [39]. Linear elastic behavior of the material is assumed for calculating the bending stresses and the elastic modulus from the unnotched cantilever bending tests using the following equations:

$$\sigma_b = \frac{6\,PL}{BW^2} \qquad \text{Eq. 1}$$

$$E = \frac{4P}{B\delta}\left(\frac{L}{W}\right)^3 \qquad \text{Eq. 2}$$

where, $P$ is the applied load, $\delta$ is the displacement of the cantilever beam at the point of loading which equals the displacement recorded by the nanoindenter.

The load values were normalized with the microcantilever dimensions using Eq. 3 to determine the stress intensity factor ($K_I$) *for* the notched cantilevers. The critical value of $K_I$ at which fracture occurs under plane strain condition is the fracture toughness ($K_{IC}$) of the material. $K_{IC}$ of the tested films was calculated from the critical load ($P_c$) at which the crack propagates *i.e.* maximum load at which fracture occurs, using Eq. 3 and Eq. 4. The notch depth $a$ was measured from the top edge of the cantilever face to the notch tip from the SEM images of the fracture surface post bending. In displacement-controlled tests, as reported here, crack propagation results in distinct load drops.

$$K_I = \frac{P\,L}{BW^{1.5}}\, f\left(\frac{a}{W}\right) \qquad \text{Eq. 3}$$

$$f\left(\frac{a}{W}\right) = -3.15 + 72.85\left(\frac{a}{W}\right) - 188.51\left(\frac{a}{W}\right)^2 + 202.61\left(\frac{a}{W}\right)^3 \qquad \text{Eq. 4}$$

where, $f(a/W)$ is a geometric factor that accounts for the geometry of the crack and fracture testing sample [38].

The dimensions of the cantilevers need to be greater than the thickness required to transmit stress in the sample in full plane strain conditions *via* relationship shown in the Eq. 5.

$$B \geq 2.5 \left(\frac{K_{Ic}}{\sigma_y}\right)^2 \qquad \textit{Eq. 5}$$

where, $\sigma_y$ is the yield stress of the material.

### 2.4 Finite element modelling

Finite element modeling (FEM) analysis was performed on the PLD a-$Al_2O_3$ cantilevers which showed bending plasticity. A three-dimensional simulation setup of cantilever bending was developed in a commercial finite element (FE) analysis software Abaqus, to model the PLD a-$Al_2O_3$ cantilever bending tests and to estimate the stresses developed in the beam. The cantilever beam was represented as a rectangular cuboid with length $L_l$ = 17 µm, thickness $W$ = 3 µm, and width $B$ = 2.5 µm as shown in Fig. S2 (supplementary information section S2). The cantilever beam is attached to another rectangular cuboidal region where the fixed boundary conditions (BCs) are applied. This cuboidal region was meshed with the cantilever in such a way that they share the same FE nodes at the common surface, realizing perfect contact conditions. Mesh element types of 8-node linear brick (C3D8) and 4-node tetrahedral (C3D4) elements were used in the FE model with an average mesh size of 0.1 µm. Amorphous $Al_2O_3$ was modelled as isotropic elastic-plastic material. Poisson's ratio was set to 0.3 and no hardening was assumed in the material. The load was applied at length $L_l$ = 14 µm to a reference point (as shown in Fig. S2), mimicking the experiments, while controlling the row of nodes via multi-point constraint on the beam. The yield stress of the material was set to 5.6 GPa, according to the micropillar compression model used in our previous study on the same material [4]. Tension-compression symmetry in both elastic and plastic response behavior is assumed for bending.

## 3 Results

### 3.1 Structural characterization and chemical composition measurements

The representative XRD spectrum of ALD, SD and PLD $Al_2O_3$ films are shown in Fig. 2a. All films were found to be amorphous and no crystalline peaks of $Al_2O_3$ were observed. A low intensity crystalline peak visible between 60 – 65º in the XRD spectra of the SD film belongs to the interlayer TiAlN used on Si substrate during deposition. The amorphous nature of all film systems was confirmed using TEM diffraction measurements. Fig. 2b-2d shows the bright field

TEM images of ALD, SD and PLD deposited $Al_2O_3$ films along with corresponding diffraction patterns in the inset. The diffused amorphous pattern confirms that all the films are completely amorphous.

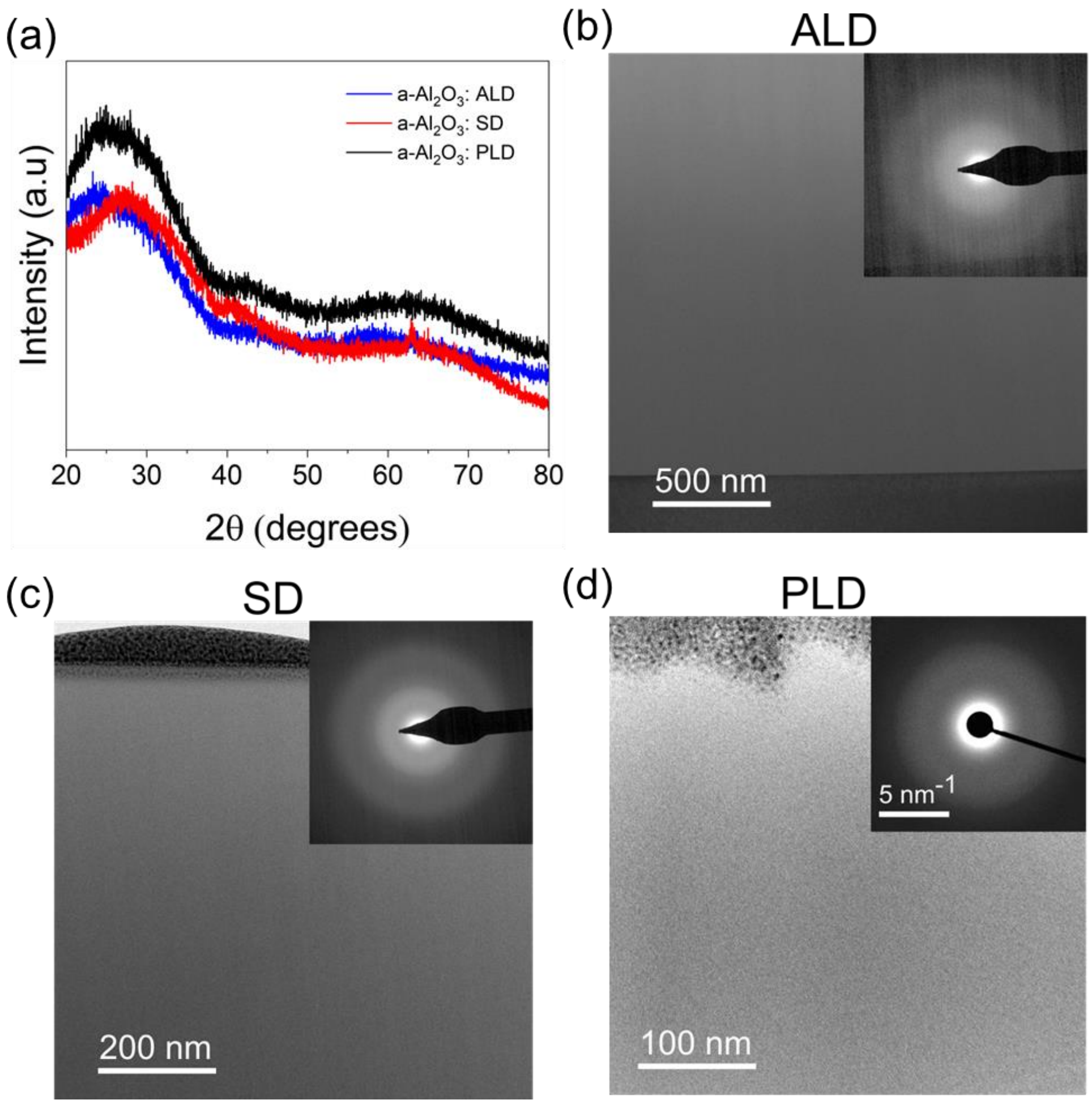

*Fig. 2: GI-XRD and TEM characterization of a-$Al_2O_3$ films. a) Representative GI-XRD spectrum a-$Al_2O_3$ film systems showing amorphous spectra. bright field TEM images and diffraction patterns(inset) of b) ALD, c) SD and d) PLD films confirming the amorphous structure.*

Preliminary observation from energy dispersive spectroscopy (EDS) showed the atomic ratio of oxygen-to-aluminum (O/Al) to be close to the expected stoichiometry for $Al_2O_3$ for all deposited films. SEM-EDS analysis from all a-$Al_2O_3$ films yielded an O/Al ratio of 1.5. Overall, samples appeared chemically homogenous with minor variation in chemical composition from repeated measurements. More detailed investigation of the stoichiometry and elemental composition was carried out using RBS and ToF-ERDA measurements. The RBS spectra from the three a-$Al_2O_3$ films, showing the Al and O yield, are shown in Fig. 3a. The respective ToF-ERDA profiles of the a-$Al_2O_3$ films are shown in Fig. 3b-3d which provide quantitative, depth-resolved, elemental compositions of Al, O and H present in the film. The depth resolved concentration profile shows similar Al and O at.% confirming the near-perfect stoichiometry of deposited films. Note that the elemental compositions were analyzed only in the depth range between the blue and red markers (Fig. 3b-3d). The O/Al atomic ratio of all the films determined from RBS and ToF-ERDA are listed in Table 1. The O/Al ratio in all three films was estimated to be closer to the theoretical value of 1.50 from both these analyses, confirming that the three amorphous films have near-perfect stoichiometry. It should be noted that an error of up to ~ 7% can be expected in the elemental ratio values obtained from ToF-ERDA analysis. The ToF-ERDA results on ALD film show the presence of impurities with < 5 at.% hydrogen and 0.009 at.% carbon. These impurities originate from the unreacted metal-organic precursor ligands or from unreacted OH groups during the sequential ALD surface reactions. SD and PLD films were found to be free of these impurities. To conclude, all three a- $Al_2O_3$ films are relatively free of chemical impurities (except for the ALD film that contains hydrogen) and exhibit near-perfect stoichiometry making them comparable from structural and chemical point of view.

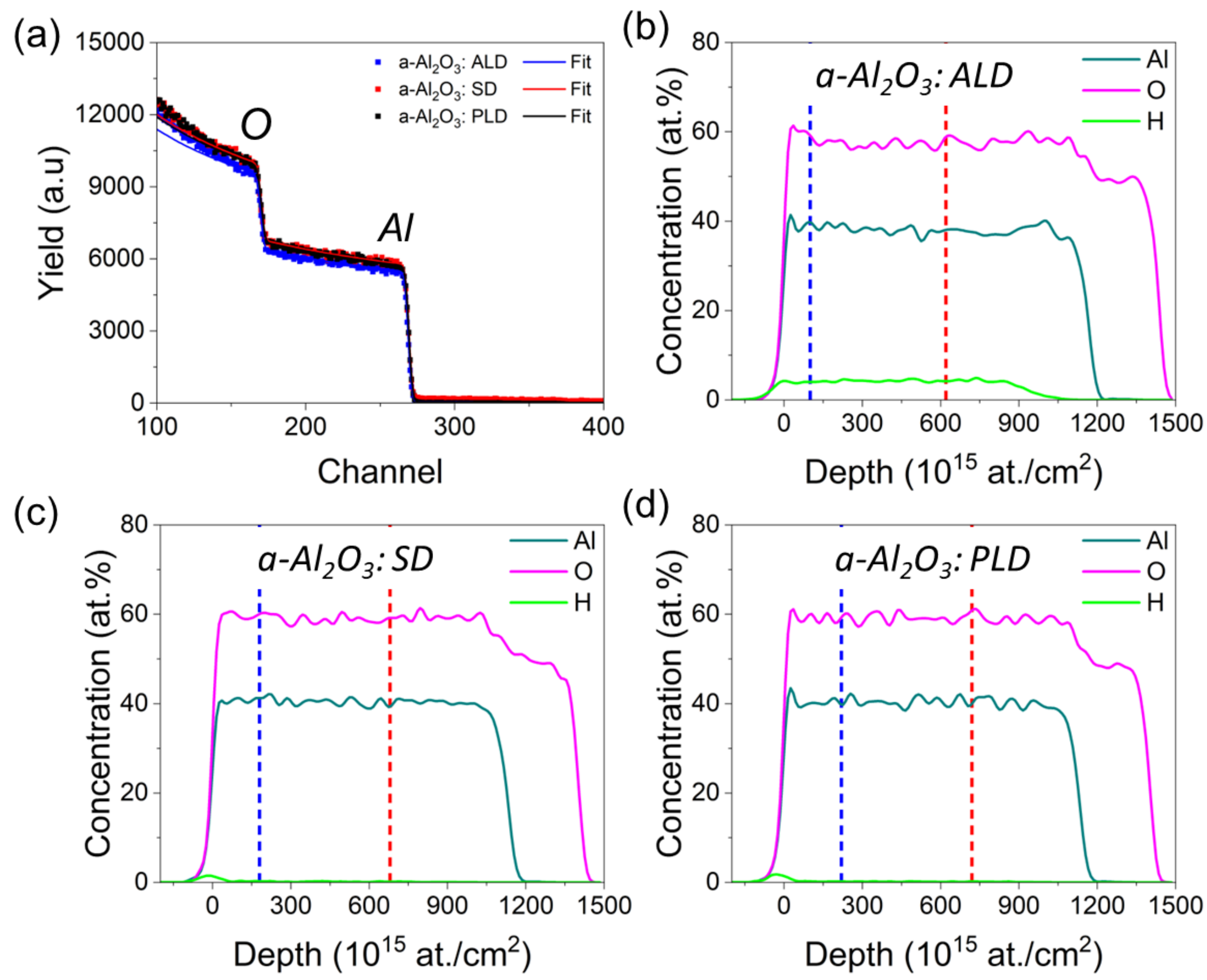

*Fig. 3: Elemental analysis of deposited a-$Al_2O_3$ films. a) RBS spectra of a-$Al_2O_3$ films confirming the composition. ToF-ERDA spectra showing the depth resolved concentration profile of b) ALD, c) SD and d) PLD a-$Al_2O_3$ films.*

*Table 1: O/Al elemental ratios of the a-$Al_2O_3$ thin film samples measured using time-of-flight elastic recoil detection analysis (ToF-ERDA) and Rutherford backscattering spectrometry (RBS).*

| ***Film type*** | ***O/Al atomic ratio*** | |
|---|---|---|
| | ***ToF-ERDA*** | ***RBS*** |
| a-$Al_2O_3$: ALD | 1.50 ± 0.1 | 1.52 ± 0.06 |
| a-$Al_2O_3$: SD | 1.48 ± 0.1 | 1.50 ± 0.06 |

| $a-Al_2O_3$: PLD | 1.48 ± 0.1 | 1.46 ± 0.06 |
|---|---|---|

### 3.2 Flexural behavior of $a-Al_2O_3$ films

The representative bending stress-strain curves for each of the $a-Al_2O_3$ films are shown in Fig 4a. The corresponding load-displacement data is included in Fig. S5a (supplementary information section S3). The flexural behaviour of the unnotched microcantilevers of $a-Al_2O_3$ films shows a slight non-linear elastic response intially. This non-linearity arises possibly due to sliding of the wedge indenter tip on the cantilever top surface. The bending data indicates a drop in contact stiffness due to sliding of the tip even in the elastic regime. Thereafter, the flexural behavior diverges based on the sample studied. Fig. S6 in supplementary information section S4 shows the contact stiffness evolution during the bending experiment.. All SD cantilevers show (nominally) linear elastic brittle failure. This failure appears as a vertical load drop to zero in our displacement-controlled tests. In contrast, all PLD cantilevers exhibit substantial plastic deformation after the initial elastic regime. At high strains, the indenter tip visibly slid towards the free end of the cantilever, from the initial loading position, and cleaved a small part of the top surface in a ductile manner (Fig. 4d). The bending test was discontinued when the indenter made contact with the FIB milled trench edges around the cantilever or when the cantilever touched the base material. Partial elastic recovery was observed upon unloading. Note that no fracture occurs at peak stress for PLD $a-Al_2O_3$ films in Fig 4a as the bending test was discontinued at that point. Also note that the bending stress and strain values reported in the plastic regime can be inaccurate because simple elastic beam bending theory was used to estimate them which is not valid for the plastic regime. Interestingly, ALD $a-Al_2O_3$ cantilevers show vastly varying deformation behavior. Two of the ALD cantilevers showed linear elastic brittle failure while the other two showed plastic deformation without failure. Two representative stress-strain curves corresponding to each deformation type are shown in Fig 4a while the load displacement curves are shown in Fig. S5b in supplementary information section S3. The fracture stress of the ALD $a-Al_2O_3$ cantilevers that failed in a brittle manner was 9.4 ± 0.4 GPa which is ~ 70% higher than that of SD $a-Al_2O_3$ cantilevers (5.5 ± 0.8 GPa). The fracture surfaces for both of these films are shown in Figs. 4b and 4c. The ALD film exhibits a smooth and flat cleavage fracture whereas the sputtered SD film shows irregular features that are indicative of columnar film growth. SD film fracture surface clearly shows processing induced flaws due to columnar grain growth.

SEM images of the PLD microcantilevers post deformation, an example shown in Fig. 4d, show clear remnant plastic deformation in agreement with the residual plastic strain observed in bending stress-strain curves. The slope of the loading and unloading segments differ due to the indenter tip sliding on the top surface and also, penetrating into the material. Both micromechanical bending data and the observed remnant plasticity post bending conclusively prove plasticity in PLD a-$Al_2O_3$ films. Due to the presence of tensile stresses at the top of the cantilever, tensile plasticity of PLD films at micrometer length scales is conclusively established through these experiments at room temperature. The stress distribution in the PLD cantilever at maximum loading condition estimated using FEM simulations is shown in Fig. 4e (also refer to Fig. S4 in the supplementary information section S2 for further details). Tensile stresses are generated above the neutral axis and closer to the fixed end of the cantilever. Maximum stress values of ~ 6.3 GPa were predicted from FEM simulations for the maximum displacement value corresponding to our microcantilever bending experiments. The maximum normal tensile stress developed in the cantilever exceeds the yield stress value of 5.6 GPa, which was determined in our previous study on PLD films using a combination of micropillar compression experiments and FEM simulations [4]. This maximum tensile stress value from FEM simulations matches well with the yield stress value of 5.1 GPa observed from the experimental bending stress data in Fig 4a. We observe a higher yield stress for the PLD films from our bending experiments than what has been reported earlier [3,4,40]. In the bending stress calculations using *Eq. 1*, we assume a linear elastic behavior which does not take into account the plasticity nor the sliding of the indenter tip. Although bending stress values for PLD film in Fig 4a seem to continuously increase after yielding, this is not a real material hardening effect but likely comes as an artefact from the combination of test setup and calculation method.

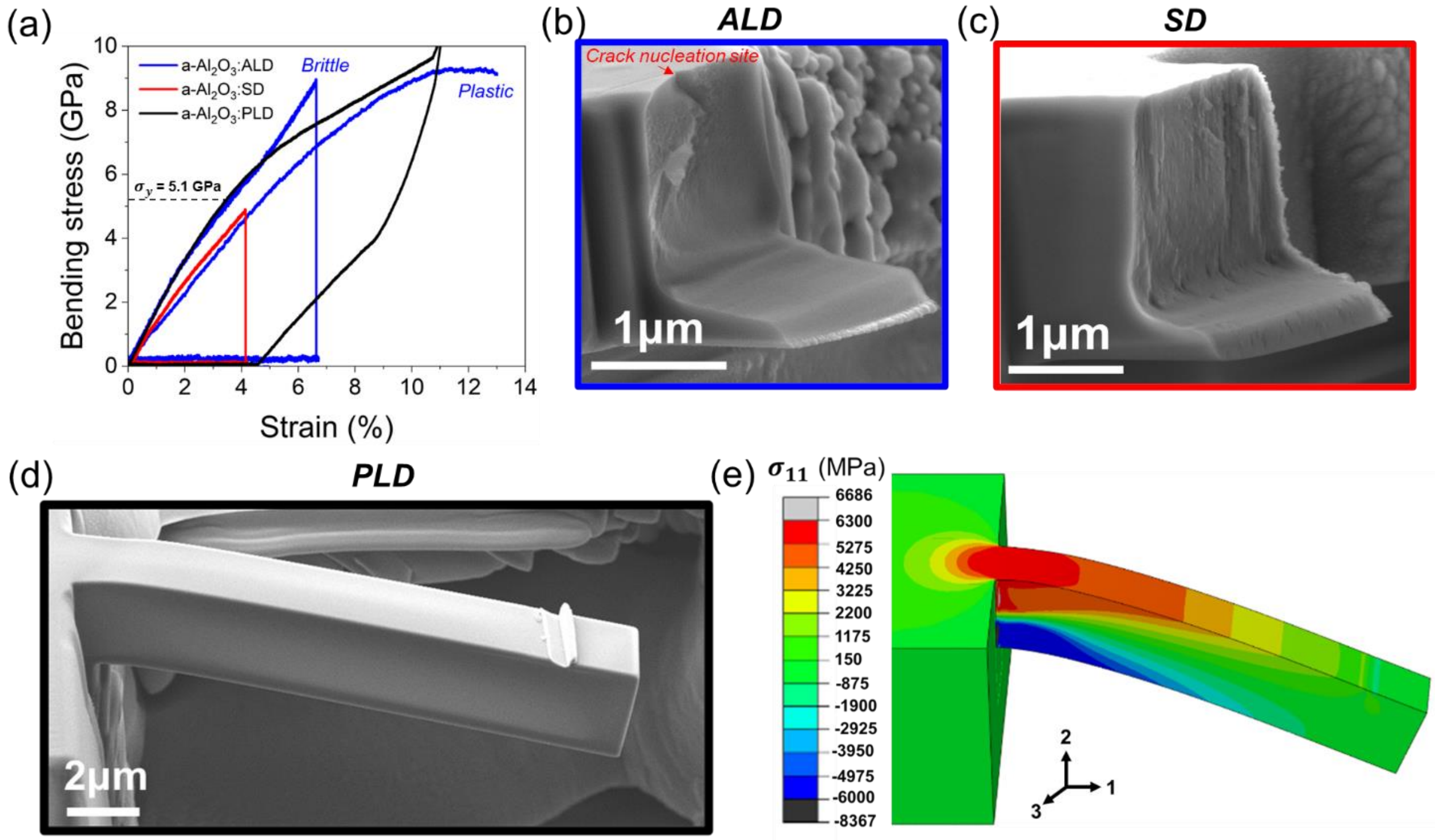

*Fig. 4: a) Representative bending stress-strain curves of a-$Al_2O_3$ films synthesized with ALD (blue, brittle and plastic behavior), SD (red) and PLD (black) obtained during the microcantilever bending test. Fracture surface images of b) ALD and c) SD a-$Al_2O_3$ cantilevers. d) Post-deformed image of a PLD a-$Al_2O_3$ microcantilever showing plastic deformation after bending. e) Normal stress distribution in the PLD a-$Al_2O_3$ microcantilever, $\sigma_{11}$ represents stresses acting along direction 1 and perpendicular to 2-3 plane) where positive and negative values indicate tensile and compressive stress, respectively.*

### 3.3 Fracture toughness of a-$Al_2O_3$ films

Notched microcantilever bend tests showed linear elastic response for all three a-$Al_2O_3$ samples followed by catastrophic failure in a brittle manner when the critical stress was reached. The representative stress intensity factors are plotted as a function of displacement for the three a-$Al_2O_3$ films in Fig. 5a, and the corresponding load-displacement data for all microcantilevers is shown in Fig. S7 (supplementary information section S5). No direct evidence of crack tip plasticity in notched cantilever fracture tests was observed in our *in situ* experiments or from the load-displacement curves, which would otherwise have shown as load drops and/or changes in the loading slope. The post fracture images of the microcantilevers for all the film systems are shown in Fig. 5b-5d. Reasonably smooth cleavage surfaces are observed for ALD and PLD films. Slight evidence of crack propagation through the columnar grain boundaries is observed for the SD film, similar to the flexure tests reported in the previous section. However, the fracture surface is much smoother this time suggesting that crack deflection along the columnar grain boundaries is minimal. These fracture surface observations are consistent with linear elastic brittle failure observed in microcantilever fracture experiments. Fracture toughness values of $3.3 \pm 0.4$ MPa.m$^{0.5}$, $3.0 \pm 0.3$ MPa.m$^{0.5}$ and $3.0 \pm 0.2$ MPa.m$^{0.5}$ were obtained for ALD, SD and PLD a-$Al_2O_3$ films respectively. Fracture toughness values of the three films were nearly identical to each other with a marginally higher value measured for the ALD thin films.

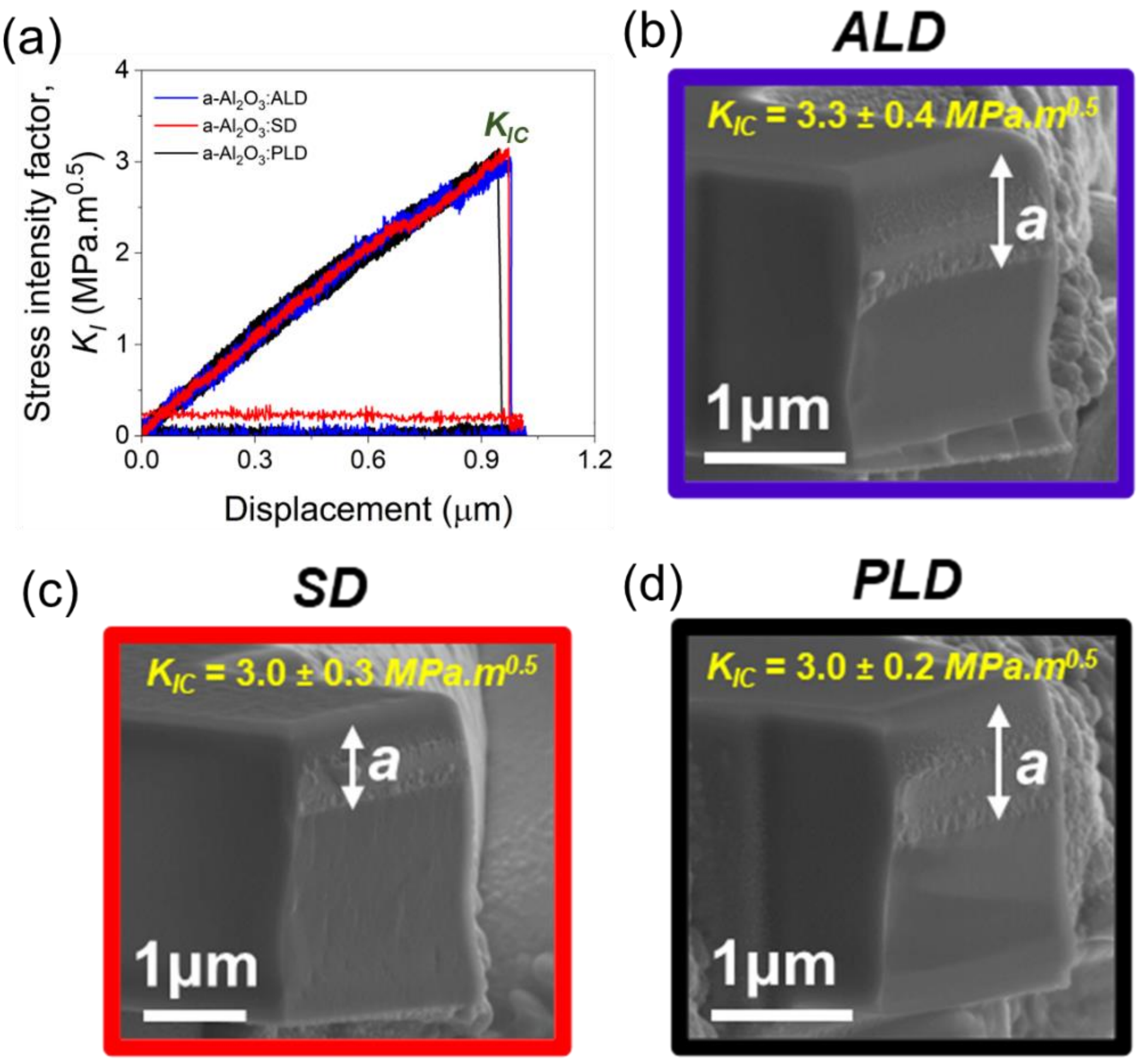

*Fig. 5: a) Stress intensity factor vs. displacement curves for ALD, SD and PLD a-$Al_2O_3$ films measured using the microcantilever fracture test. Microcantilever fracture surface images of b) ALD, c) SD and d) PLD a-$Al_2O_3$ films, where 'a' denotes the initial notch depth.*

### 3.4 Nanoindentation properties of a-$Al_2O_3$ films

Nanoindentation hardness and modulus results obtained from the three a-$Al_2O_3$ films are summarized in Table 2, along with microcantilever bending and fracture results, for comparison. These measured hardnesses and elastic modulii are comparable to previous results in literature on a-$Al_2O_3$ which report 8 - 11 GPa hardness and 169 - 200 GPa elastic modulus [41–47]. The nanoindentation experiments on these films estimate a hardness value ~ 10 GPa for all a-$Al_2O_3$ films, whereas the elastic modulus of PLD film is approximately 9% higher than that of ALD and SD films. This suggests that the PLD films may have slightly higher density compared to the other two films, which would also result in less defects that can act as stress concentrators aggravating the failure.

Based on the current results, the elastic modulus estimated from the microcantilever bending tests is lower than nanoindentation modulus by ~ 20-35%. This discrepancy can arise due to difference in nature of the stress states between both the tests and potential densification under the hydrostatic stress beneath the indenter tip. The confined hydrostatic stress component in nanoindentation can result in further densification leading to increased indentation modulus. A densification of ~ 3% in a-$Al_2O_3$ was reported by Frankberg et al. [4] during compression using molecular dynamics simulations. Another reason for this difference arises from the δ (i.e. displacement) value used in Eq. 2 for calculating the elastic modulus from bending experiments. This δ comprises of two components: (a) the displacement due to beam bending itself, and (b) indentation of the tip into the cantilever *i.e.* indenter penetration into the material. The additional tip penetration overestimates the δ value, which in effect lowers the elastic modulus determined from cantilever bending tests, leading to discrepancy with nanoindentation modulus estimates. The contact stiffness evolves with displacement due to tip penetration into the material, and surface sliding during the bending experiment, which influence the modulus estimates from bending tests (supplementary information section S4).

*Table 2: Summary of the mechanical properties of the different a-$Al_2O_3$ thin film systems*

| ***Film type*** | ***Hardness, H (GPa)*** | ***Elastic modulus, E (GPa)*** | | ***Maximum flexural strength, $\sigma_b$ (GPa)*** | ***Fracture toughness, $K_{IC}$ ($MPa.m^{0.5}$)*** |
|---|---|---|---|---|---|
| | | ***from nanoindentation*** | ***from bending*** | | |
| a-$Al_2O_3$: ALD | 10.7 ± 0.4 | 175 ± 8 | 146 ± 7 | 9.4 ± 0.4 | 3.3 ± 0.4 |
| a-$Al_2O_3$: SD | 10.1 ± 0.4 | 179 ± 6 | 137 ± 10 | 5.5 ± 0.8 | 3.0 ± 0.3 |
| a-$Al_2O_3$: PLD | 10.4 ± 0.2 | 190 ± 7 | 139 ± 12 | 9.5 ± 0.9 | 3.0 ± 0.2 |

# 4 Discussion

SEM images of FIB fabricated microcantilevers and cross-sections from milled trenches did not show any visible pores or defects during cantilever preparation, suggesting that relatively dense films were obtained from all three deposition methods (Fig. 1). Density of films vary in relation to synthesis conditions, densities $\geq$ 3.45 g/cm$^3$ have been observed with PLD film grown at room temperature [46], whereas densities |> 3.0 g/cm$^3$ have been reported for ALD films for growth temperature above 120°C [48–50]. This is consistent with previous literature reports that show that all three deposition methods are capable of producing $Al_2O_3$ films with relative densities above 75% compared to density of crystalline $Al_2O_3$ (3.98 g/cm$^3$). Even though we have not performed any quantitative density measurements on these a-$Al_2O_3$ films, an initial estimate of density can be inferred from the nanoindentation elastic modulus values. PLD film, which has the highest modulus and is chemically pure, should have the highest density among these films. Slightly lower indentation modulus observed for the ALD film, compared to the PLD film, is likely due to the addition of significant amounts of low density impurities consisting of hydrogen/hydroxide and carbon/hydrocarbon/carbon oxide species [51,52]. In this study, the ALD film contained less than 5 at.% of H and 1 at.% of C which lower the elastic modulus of the ALD film [53]. The SD film, though free of such chemical impurities, shows comparable modulus to the ALD films suggesting the presence of defects. Extensive columnar growth was observed in the SD film, resulting in column boundaries or intercolumnar voids, which decreases the overall film density. Consequently, the ALD film can be considered to be more homogeneous in comparison to SD film.

The PLD a-$Al_2O_3$ microcantilevers exhibit extensive plastic deformation in bending, with yield stress of ~5-6 GPa consistent with our previous micropillar compression study [4]. All SD cantilevers fractured at 5 GPa stress, whereas half of ALD cantilevers showing brittle behavior failed at ~9 GPa. This fracture strength is still higher than the previously reported fracture strength of ALD films of ~3 GPa from bulge tests [54], but well below the theoretical strength of a-$Al_2O_3$ of ~12.5 GPa (estimated from $G/10$, with shear modulus, $G$ = 125 GPa [55]). The fracture of ALD alumina microcantilevers observed in this study could possibly initiate from processing flaws in the deforming region. The crack initiated from an existing flaw subsequently extends along the cross-section and culminates in the final fracture of the microcantilever beam. We notice additional crack fronts on the fracture surface which could have initiated from existing microscopic flaws located in the fixed end. The other microcantilevers, that did not show fracture, probably did not have processing flaws of critical size in the highly stressed region close to the fixed end. This suggests that the hydroxyl OH groups do not hinder the plastic deformation mechanism in ALD a-$Al_2O_3$ films when the films are sufficiently defect free. Recent experimentally curated simulations [56,57] show that ALD a-$Al_2O_3$ consists of several nearest neighbor coordination sphere (NNCS) polymorphs which can be described generically as $[Al(O)_{n-m}(OH)_m]$ (with $n, m \epsilon \mathbb{N}$, $n \geq m$, $4 \leq n \leq 6$) depending on the unreacted hydroxyl groups forming during ALD at a given temperature. For 120°C ALD the polymorphs are distributed as ≈ 5% of $Al(O)_6$, ≈ 30% of $Al(O)_5$ and $Al(O)_4$ each, and ≈ 10% of $AlO_4(OH)$ and $AlO_3(OH)$ each; with a few less frequent polymorph types adding up to 100%. The distribution of these NNCS polymorphs results in the reduced density of ALD alumina films observed. The simulations also suggest water void formation when post-annealing the material starting at around 577°C (thermal energy $kT$ = 74 meV). Such energies can be transferred during transmission or scanning electron microscope (TEM and SEM) observation in knock-on collisions [58]. According to mass and kinetic moment conversation the maximum transferable energy $E_{max}$ an electron with mass $m_e$ and energy $E_0$ can give to an atom of mass $m_a$ is $E_{max} \cong 4E_0 \cdot m_a/m_e$ (non-relativistic approximation). Setting $E_{max} = kT$ we arrive at 0.5 keV and 0.04 keV for an electron to transfer 74 meV to an oxygen and hydrogen atom, respectively. Typical observation energies with SEM are 1 to 30 keV and for TEM around 100 to 300 keV, both observation energy ranges thus being able supplying enough energy to potentially introduce voids into the ALD material. Bending plasticity observed in all PLD microcantilevers and half of the ALD cantilevers tested is likely due to extensive bond switching activities that require a sufficiently dense and 'flaw free' glassy microstructure to operate below the critical stress intensity [3]. Previous atomistic simulation studies show that the plasticity in amorphous oxides such as a-$Al_2O_3$ and amorphous $SiO_2$ (a-$SiO_2$) occur primarily through cation-oxygen bond switching in the short atomic range [3]. Further, in the short-to-medium atomic range, amorphous oxides form cation-centered polyhedral structures. The plasticity in a-$Al_2O_3$ and a-$SiO_2$ has been shown to occur by polyhedral neighbor change events (PNCE) between adjacent polyhedra, where edge-sharing polyhedra have the most significant contribution to plasticity [35]. For a detailed understanding of bond switching activities responsible for plasticity in amorphous alumina, the reader is referred to these studies [3,4]. In comparison, both ALD and PLD exhibit very smooth fracture surfaces (Fig. 4b, 5b and 5d). There were no visible shear bands on the surface of any of the plastically deformed cantilevers, similar to our previous observation on compressed micropillars at lowest tested strain rates ($10^{-3}$ $s^{-1}$) [4]. The reason for this can be the used quasi-static strain rate of deformation in bending which allows homogenous deformation throughout the deformation volume and prevents any strain localization. In contrast to plastic behavior, the SD microcantilevers fail catastrophically beyond the elastic limit likely due to defects observed at the column boundaries that act as stress concentrators. The fractured cross-sections of SD a-$Al_2O_3$ observed in both notched and unnotched cantilevers reveal that the fracture surface has rough irregularities (Fig. 4c and 5c). The vertical alignment of these irregularities, parallel to the growth direction, indicates that SD deposition induced material defects related to the columnar growth that act as weak links leading to fracture without any plastic deformation. Cross-section analysis of alumina films in SEM did not reveal any microscale defects in the films as all of them appeared completely dense (Fig. S8 in supplementary information section S6). Linear features of 1-2 nm in width are observed in the SD a-$Al_2O_3$ as bright contrast in the bright-field TEM image due to mass-thickness contrast, indicating regions of reduced electron scattering (Fig. S9 in supplementary information section S6). These same features are not visible in the corresponding dark-field image, which rules out the possibility of crystalline inclusions or oriented nanocrystalline phases that would otherwise produce diffraction contrast. The absence of dark-field signals in an otherwise amorphous matrix is consistent with these features being true voids or nanoscale column boundaries. The linear morphology and alignment of these features along the film growth direction further supports their identification as inter-columnar voids formed during SD deposition, a characteristic signature of the columnar microstructure evolution. The effect of hydrogen on plasticity in amorphous films is not conclusively established. The hydroxy bonds formed in the ALD deposited films can potentially act as weak links for cracks to propagate easily and promote brittle fracture. This influence of hydrogen rich regions was speculated to promote fracture in amorphous silicon oxide films in bending previously [59]. However, amorphous silicon oxide also has been shown to have more atomistic level cavities through atomistic simulations, which inhibit bond switching [60]. In our study, the extensive plastic deformation observed in the ALD microcantilevers, which did not fracture, suggests that a-$Al_2O_3$ films can exhibit plasticity in spite of the presence of H impurities. The critical condition seems to be having a dense enough structure to enable sufficient bond switching to accommodate the plastic strain. This is extremely encouraging from application point-of-view because obtaining ultra-pure amorphous oxide films is not necessary. Up till this point, thick ALD alumina films (≥ 5 nm, H content 10-30% [61–63]) have generally been considered to be brittle. This stems from lack of adequate mechanical characterization of the ALD films using direct testing methods. We show, for the first time, that ALD films can indeed show plasticity with ~ 2 µm cross-section microcantilevers. We must point out that milling these cantilevers reliefs the residual stresses generated during film growth, thereby isolating the intrinsic ductile behavior. If such thick films can exhibit plasticity, it is highly likely that 10's of nm thick alumina films typically used in semiconductor industry and multi-layer research should show plastic deformation as well.

The remarkable tensile plasticity observed in the PLD films during unnotched microcantilever bending seems to have negligible contribution during the notched microcantilever fracture tests, as notched cantilevers of all three films fractured in a brittle manner, with a similar fracture toughness value. The common brittle fracture in these films, including that for PLD alumina, points towards the dominant effect of the notch, which can be considered to be a large defect promoting brittle failure. This clearly shows that the plastic deformation mechanisms observed in the unnotched bending experiments are insufficient to retard the crack propagation, as typically observed in ductile metals. The bond switching mechanism, by itself, is not capable of introducing significant crack tip plasticity. The measured $K_{IC}$ values for our a-$Al_2O_3$ films from notched microcantilever bending tests are comparable to that reported for crystalline α-$Al_2O_3$ in recent studies (2.2 - 3.7 $MPa.m^{0.5}$) [30,31], but much higher than the $K_{IC}$ for ALD a-$Al_2O_3$ (1.3 - 1.9 $MPa.m^{0.5}$) reported from shear lag tensile tests [33,64]. Shear lag test results on films attached to the substrate can provide fundamentally different results when compared to our notched cantilever bending fracture tests which consider only the film and avoid any substrate effects. In accordance with similar fracture toughness across samples, there was no evidence of any substantial crack tip plasticity around the notch tip, which could have modified the brittle failure of the material during the fracture tests. So, the presence of flaws in these films forces the material to establish their inherent flaw sensitive brittle nature and $K_{IC}$ values reported for crystalline $Al_2O_3$ (~3.0 $MPa.m^{0.5}$) [30] are not similar to the values obtained from our experiments. The measured fracture toughness values and the maximum bending stresses in ALD and SD samples, yield a critical surface flaw size ($c_{crit}$) of ~24 nm and ~57 nm, respectively, when assuming a shape factor of 1. This indicates that the PLD deposited samples exhibit a process-induced surface flaw size less than 24 nm and consequently, the PLD microcantilevers show a very high degree of plasticity. The successful linking of fracture mechanics and materials synthesis shows that a manufacturing path that enables fabricating adequately flaw free structure, like PLD in this case, can be used to scale plasticity to larger length scales and thicker samples. The microcantilever bending and fracture experiments isolate the film from the substrate, thus making it stress free and delivers the true intrinsic properties. One can argue that for the already versatile family of ceramic materials, ductility would be the ultimate ceramic functionality. No prior demonstrations of such bending plasticity at room temperature exist in ceramic or inorganic glass literature to our knowledge.

While this study demonstrates bending plasticity in a sufficiently flaw-free a-$Al_2O_3$, it also reveals that the critical size flaws are in the order of 10's of nanometers and points out, that in order to obtain low temperature plasticity, ALD and sputtering techniques require further optimization to increase the film quality. Such nanoscale defects are not easily detectable, for example by SEM imaging. While TEM is able to capture these defects to some extent, it has a severe intrinsic limitation in the sample areas/volumes being probed (supplementary information section S6). Allowing bulk plasticity and aiming for the required improvements in the synthesis techniques would require characterizing the whole sample volume and therefore, new strategies to characterize nanoscale defects with larger sampling are needed. X-ray based techniques such as small angle X-ray scattering could yield information on nanoscale defects such as pores of around 1-100 nm in diameter. In addition, glassy oxides are typically transparent to visible light and recent advances in super-resolution microscopy could, in principle, allow studying defects down to the nanoscale. PLD and SD techniques are more industrially oriented and can be upscaled for production of thicker films. ALD, while limited in deposition rates, offers the possibility of obtaining conformal and large-area films on "non-flat" substrates and irregular shapes. It is extensively used in semiconductor industry and the discovery and demonstration of plasticity in ALD films potentially opens up its adoption in wider range of applications, without having to worry about nano/microscale brittle failure. Microscale plasticity in amorphous alumina opens up exciting possibilities for its usage in flexible and wearable electronics, in extreme environment conditions, H-barrier coatings and other applications where it is currently limited due to its inherent brittleness.

# 5 Note on possible error sources

On the note of addressing possible error sources, despite 5 at.% OH and similar Al:O stoichiometry all tested a-$Al_2O_3$ films the results were similar. In addition, we measure similar hardness and elastic moduli values for all films. The XRD and TEM studies confirmed that all the $Al_2O_3$ samples are amorphous making the comparison between the materials valid. The first potential source of error can arise from the use of Ga+ ions for preparing the microcantilevers and milling the notch with the FIB. The redeposition of the milled material at the notch tip can make the notch blunt, potentially shifting the fracture toughness values to higher side [65]. The dissolution of Ga+ ions and potential localized segregation at defect sites can alter the mechanical properties masking the true fracture toughness trends among the three films, if any. This can be the reason for observing similar fracture toughness values for all three film types. An alternative to preventing this effect can be using He and other ions for notching [66]. A related error from notching can also arise from the residual stresses generated at the notch tip. Previous study by Norton *et al.* [30] on notched sapphire cantilevers reported unusually high compressive stresses of ~16 GPa at the notch tip. They measured this from the change in curvature of a lamella lifted out at the notch tip. However, in this study, they did not account for the additional residual stresses due to Pt deposition prior to the lift-out, which may result in an overestimation. Nevertheless, the additional compressive residual stress due to FIB notching can potentially increase the measured fracture toughness values reported in this study. The second potential source of error can arise from the use of electron beam in our *in situ* measurements. Previous studies in the TEM, which use much higher acceleration voltages, have shown a marked change in deformation behavior due to electron exposure. Vogl et al. performed similar experiments in ALD a-$Al_2O_3$ films deposited on copper nanowires which reported highly ductile behavior up to 188% strains [67]. This study concluded that electron beam irradiation activates the bond switching mechanism within the amorphous network and promotes plasticity and underscores the effect of electron beam exposure on the mechanical properties. Luo et al. also showed the effect of electron beam irradiation on the ductility of $SiO_2$ glass nanofibers [19]. All these studies were conducted on samples with size in the order of 10s of nanometer which are irradiated with accelerating voltage of 200-300 kV, where the electron beam irradiation can influence the deformation mechanisms. In addition to the accelerating voltage, the chosen electron flux subjected to the sample area, and the resulting electron dose, has large effect on the electron beam induced dynamics. A previous study by Frankberg et al. reported *in situ* mechanical tests in TEM using 300 kV accelerating voltage, on very similar PLD alumina film as used in this study, which did not show any change in deformation behavior between beam ON and OFF conditions, as along as the tests are performed using low-intensity beam, resulting in low electron flux during the tests [3]. In addition, indicating the metastability of the films a-$Al_2O_3$, increasing the electron dose eventually leads to crystallization of the a-$Al_2O_3$ film [68]. We have observed similar crystallization of the amorphous PLD alumina film under 200 kV in TEM (Fig. S10 in supplementary information section S7). This makes it challenging to study this material in TEM. However, the *in situ* SEM micromechanical results reported here, were all performed at low voltage SEM imaging at 5 kV, keeping the acceleration voltage and imaging conditions constant for all tested samples. In one of our previous studies, we investigated micropillar compression of PLD alumina under both beam ON and beam OFF conditions, finding no noticeable change in their deformation behavior [4]. To verify such conditions in the current study, we performed similar tests on ALD, SD and PLD a-$Al_2O_3$ micropillar samples under beam ON and beam OFF conditions, and no noticeable difference in the deformation behavior was observed between beam ON and OFF conditions (Fig. S11 in supplementary information section S8). Any future study should make sure to completely preclude the influence of low kV electron beam imaging in SEM while performing *in situ* micro bending and fracture tests. In the current study, the observation of both ductility and fracture in a-$Al_2O_3$ film types, while keeping the imaging conditions constant suggests that the electron beam is not inducing the observed plasticity.

# 6 Conclusions

Deformation and fracture behavior of a-$Al_2O_3$ films deposited with three different synthesis routes (PLD, ALD and SD) were studied in microscale bending at room temperature. We report, for the first time, remarkable bending plasticity of a-$Al_2O_3$ which proves the presence of tensile plasticity at micrometer length scales with yield strength of ~ 6 GPa. Both PLD and ALD a-$Al_2O_3$ showed plasticity in microcantilever bending, whereas SD microcantilevers failed in an elastic brittle manner. Results indicate that PLD and ALD enabled the synthesis of sufficiently flaw-free a-$Al_2O_3$ which is critical for obtaining micrometer thick films for applications benefitting from damage tolerant oxide ceramic films. ALD films exhibited 70% higher bending stress than SD films but showed an elastic brittle failure at ~9 GPa in some of the tested samples. In effect, we observed a complete linear elastic brittle fracture in the SD a-$Al_2O_3$ samples and fully plastic behavior in the PLD a-$Al_2O_3$, whereas the fracture possibility of ALD a-$Al_2O_3$ seemed to depend on the criticality connected to the to presence of processing flaws in the films. The presence of defects in the material does not favor the plastic deforming ability as no crack tip plasticity was observed from the notched microcantilever tests resulting in linear elastic brittle failure in all of the notched samples. The fracture toughness values of samples from all deposition methods were found to have a similar value of $\sim 3.1 \pm 0.2$ MPa.m$^{0.5}$. However, it remains possible that these methods can be tuned to fabricate films with lower defect density with the aim of increasing their damage tolerance. This paves the way for its adoption in technologies that require functional amorphous oxides films in the nanometer to micrometer length scales with better structural integrity. To achieve improved damage tolerance at the application level substantial further research on processing of flaw-free amorphous oxide films is required.

## Author contributions: CRediT

**Nidhin George Mathews:** Conceptualization, Data curation, Formal Analysis, Investigation, Methodology, Visualization, Writing – original draft, Writing – review & editing. **Erkka J. Frankberg:** Conceptualization, Formal Analysis, Visualization, Funding acquisition, Writing – original draft, Writing – review & editing. **Vivek Devulapalli:** Investigation, Methodology, Writing – review & editing. **Chandan Kumar:** Investigation, Methodology, Writing – review & editing. **Barbara Putz:** Funding acquisition, Methodology, Writing – review & editing. **Aloshious Lambai:** Methodology. **Sergei Khakalo:** Formal Analysis, Methodology, Data curation, Writing – review & editing. **Mattia Cabrioli:** Methodology. **Bjarke Holl Christensen:** Methodology, Writing – review & editing. **Janne-Petteri Niemelä:** Methodology, Writing – review & editing. **Arnold Milenko Müller:** Formal Analysis, Methodology, Writing – review & editing. **Fabio Di Fonzo:** Methodology, Writing – review & editing. **Ivo Utke:** Methodology, Writing – review & editing. **Erkki Levänen:** Funding acquisition, Writing – review & editing. **Gaurav Mohanty:** Conceptualization, Formal Analysis, Supervision, Funding acquisition, Writing – original draft, Writing – review & editing.

## Declaration of Competing Interest

The authors declare that they have no known competing financial interests or personal relationships that could have appeared to influence the work reported in this paper.

## Acknowledgements

Authors are grateful to Leo Hyvärinen, Tampere University for help in XRD measurements, Kimmo Kaunisto, VTT Finland for valuable discussions. N.G.M, E.J.F, E.L and G.M acknowledge the use of funding from Academy of Finland grants (341050, 315451, 360436 and 338750). B.P acknowledges funding from the Swiss National Science Foundation under the Ambizione grant agreement No PZ00P2_202089. This work made use of Tampere Microscopy Center and H2MIRI facilities at Tampere University.

## Data availability

Data will be made available on request.

# Supplementary information

## S1. Nanoindentation of a-$Al_2O_3$ films

The representative load-displacement curves of a-$Al_2O_3$ films and the images of the indents from nanoindentation experiments are shown in Fig. S1. Total 25 indentations were performed on each film with a Berkovich indenter tip up to a maximum indentation depth of ~230 nm in displacement-controlled mode. The maximum indentation depths were selected to be within ~ 1/10$^{th}$ of the film thickness to avoid the elastic substrate effects during the measurement of indentation modulus. The nanoindentation curves of all films were observed to be similar. No load drops were observed in the load-displacement curves which are typical signatures of cracking during indentation on brittle materials. No  clear evidence of crack formation in the films post indentation was observed from secondary electron (SE) images, shown in Fig. S1. While some vein-like features can be observed inside the indent on SD film, it seems more likely due to compression of the top surface morphological features rather than cracks. The top surface of the PLD and ALD films are rather smooth, as seen from SE images, while SD films show rough morphology.

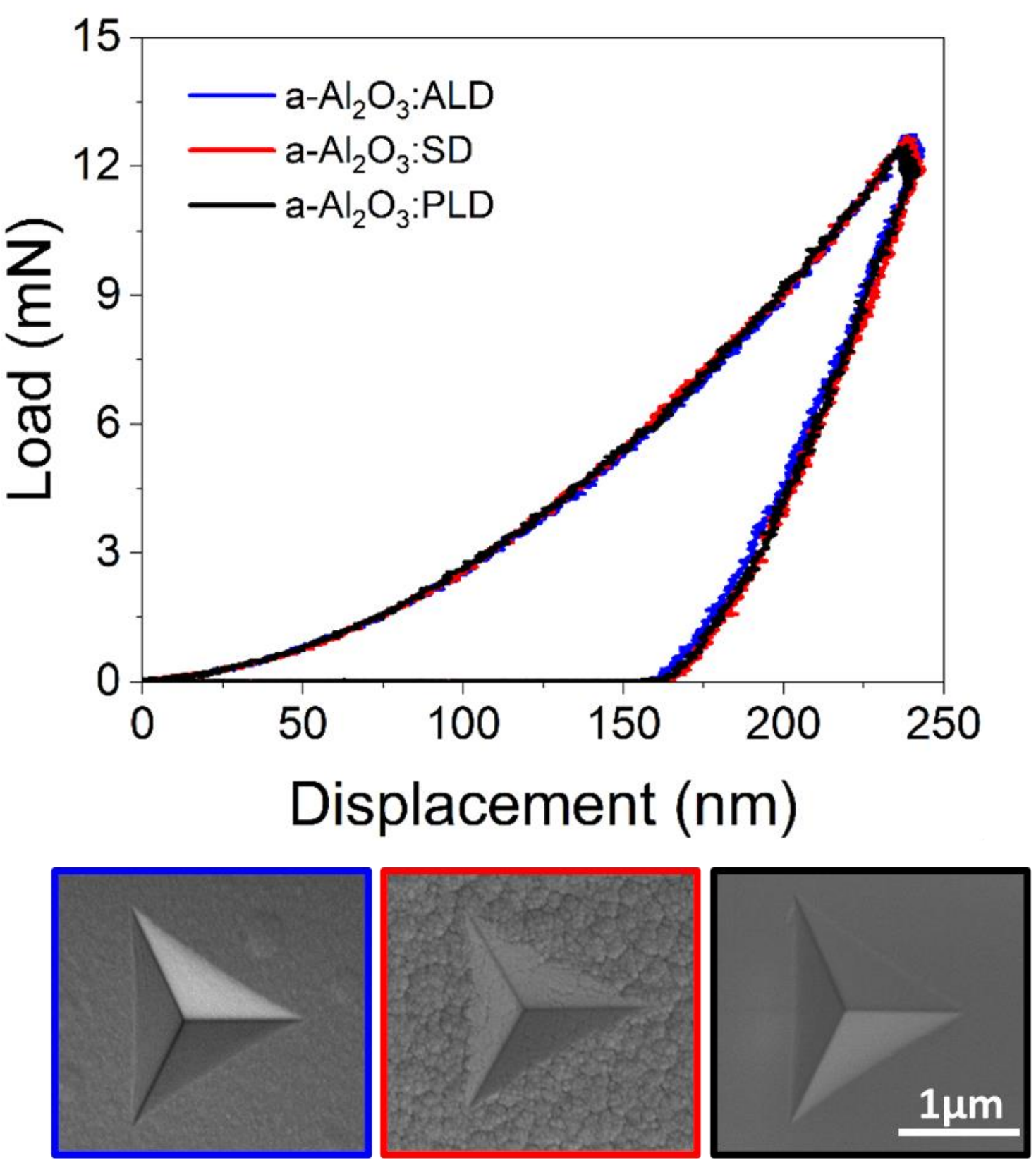

*Fig. S1: Representative load-displacement curves from nanoindentation on ALD (blue), SD (red) and PLD (black) a-$Al_2O_3$ films along with SE images of the indents showing no clear evidence of cracking.*

## S2. Finite element modelling of bending of PLD alumina cantilevers

The three-dimensional finite element simulation model of the cantilever used for the analysis is shown in Fig. S2.

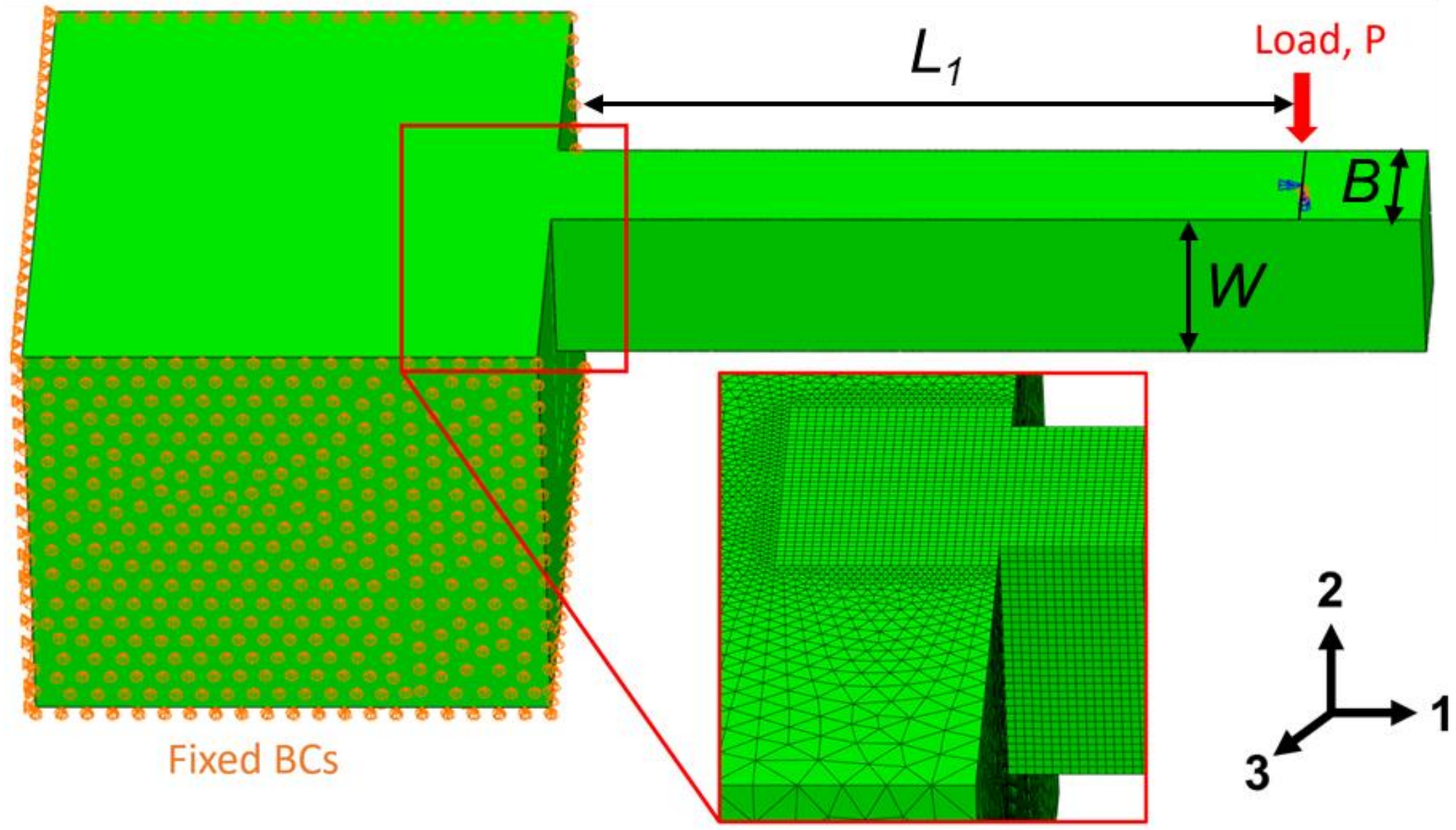

*Fig. S2: Finite element simulation setup for a PLD cantilever bending. Inset shows the type of meshing and mesh compatibility of the model.*

The elastic modulus of PLD films determined from nanoindentation ($E$ = 190 GPa, see Table 2 in main text) was used as the input describing the elastic properties of the material. Initial validation of the finite element (FE) model was performed by comparing the load-displacement data determined from the model with the actual experimental data. As shown in Fig. S3, the model with $E$ = 190 GPa underestimates the stiffness of the material obtained from microcantilever bending experiments. In order to have a better fit with the initial linear part of the experimental loading curve, the elastic modulus is considered with a value of 230 GPa. The FE model predicts linear elastic behavior prior yielding point and further hardening-like transition to a plateau in load. The simulated unloading curves are straight lines with slopes defined by the material elastic modulus. The actual loading-unloading is performed with a sharp diamond wedge tip that along with global bending deformations causes local deformations induced by the indentation and consequent scratching of the beam surface. Besides the material internal deformation mechanisms, this is believed to be the main reason behind the observed two-stage (first nonlinear and second linear) unloading. To accurately reproduce the experimental bending response, the current FE as well as material model should be developed further, which is left for future studies. The maximum stresses developed on top of the cantilever surpass the material's yield strength as seen from the distributions of von Mises and axial stress fields at maximum load in Fig. S4.

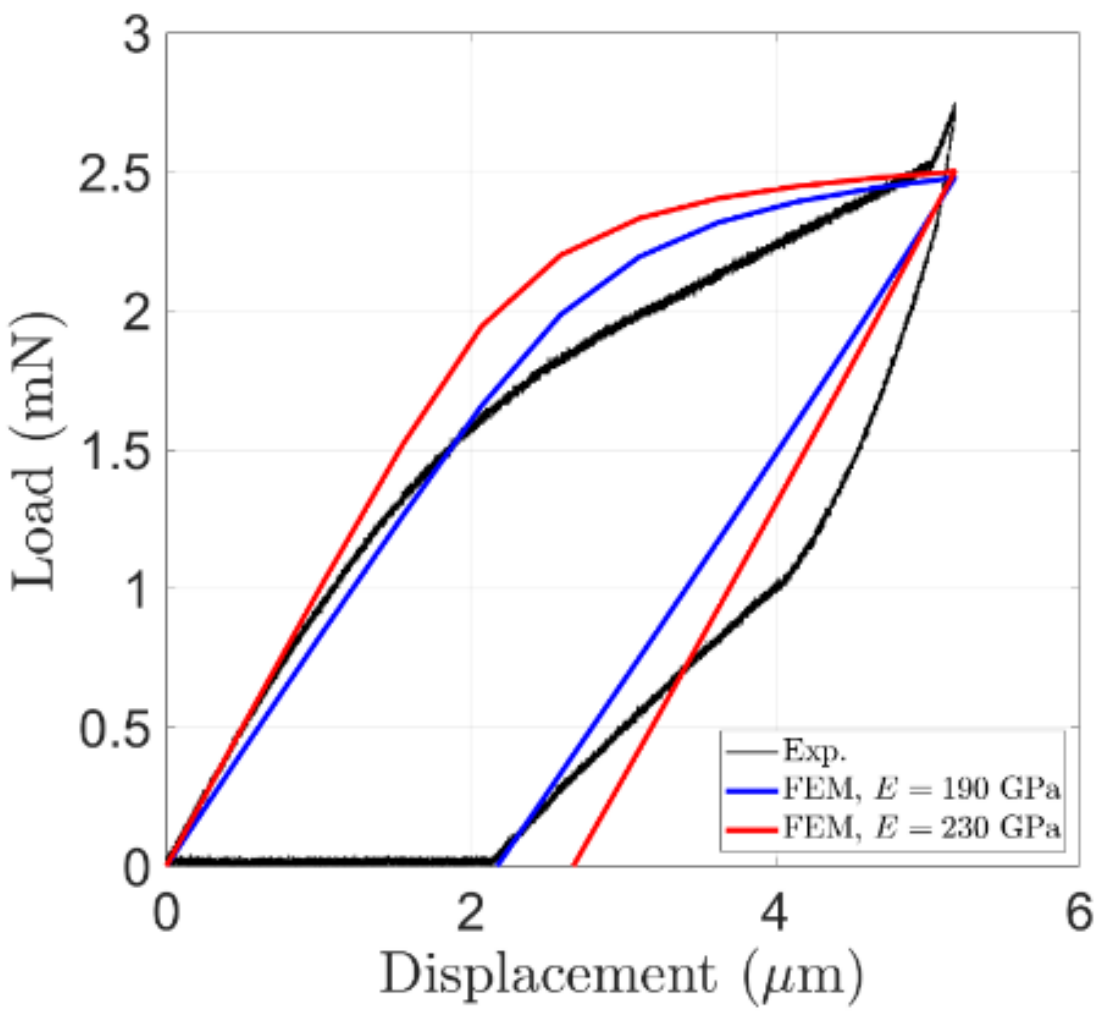

*Fig. S3: Experimental and simulated load-displacement curves of PLD a-$Al_2O_3$ films during microcantilever bending tests.*

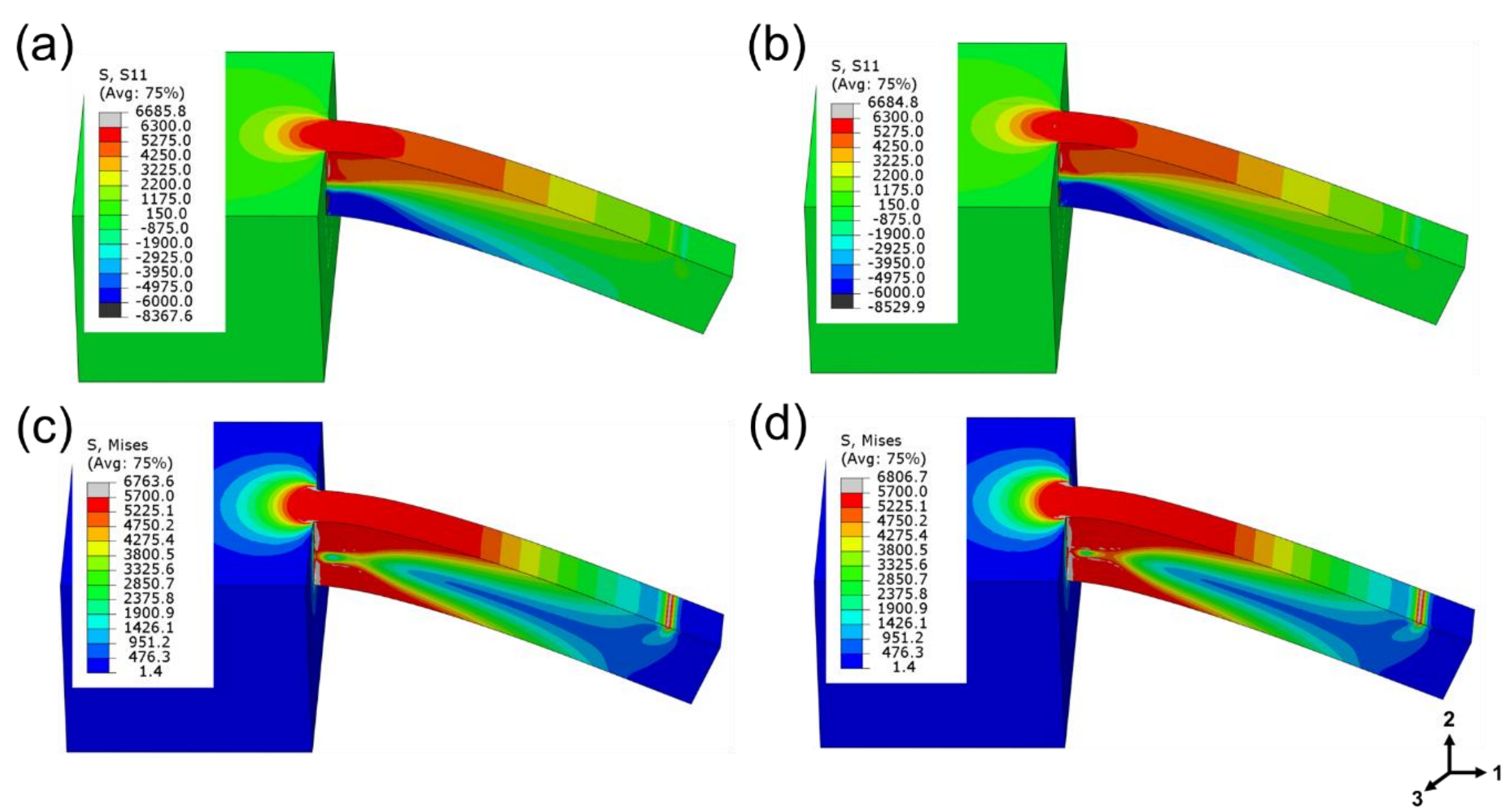

*Fig. S4: a), b) Distribution of axial (1-1 component) stress field and c), d) von Mises stress field for E=190 GPa (left) and E=230 GPa (right) at maximum loaded state.*

**S3. Microcantilever bending experiments of a-$Al_2O_3$ films**

The load-displacement response from the microcantilever bending experiments on all three alumina films are shown in Fig S5 a. The ALD and SD film showed a linear elastic brittle response before failure, whereas the PLD films showed a nonlinear response. All of the tested PLD alumina cantilevers showed plastic deformation. The drastic increase in load at maximum displacement observed for PLD cantilever is due to one of the following two reasons: (a) either the wedge indenter tip makes contact with the sides of the milled region (i.e. surrounding material), or (b) the cantilever touches the bottom of the milled region. The indenter was retracted when such large increase in loads were observed. Half of the tested ALD cantilevers failed in a brittle manner and while the rest showed plastic response in bending, as seen in Fig S5 b. All SD cantilevers failed in a brittle manner.

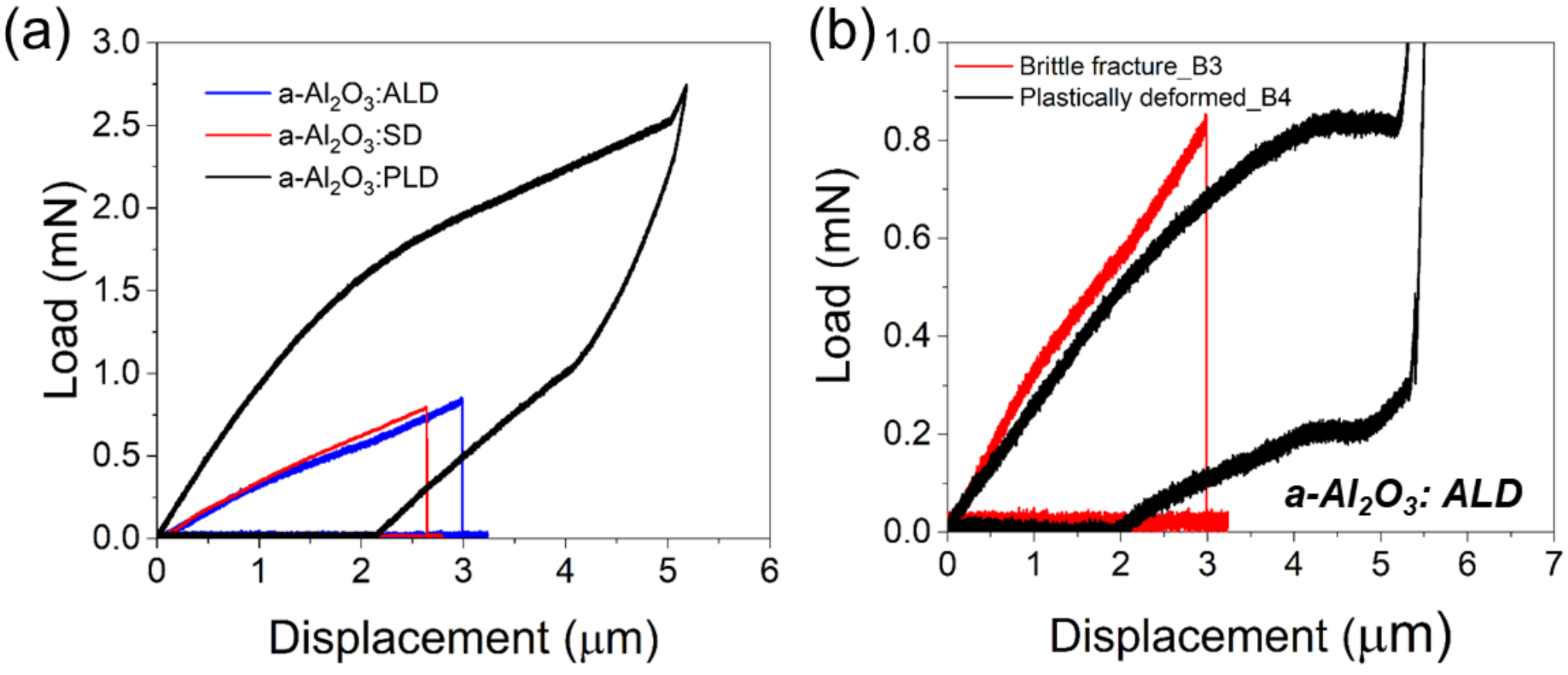

*Fig. S5: a) Representative load-displacement curves from microcantilever bending experiments performed on ALD, SD and PLD a-$Al_2O_3$ films. b) Load-displacement curve of a brittle ALD cantilever and a plastically deforming ALD cantilever.*

**S4. Contact stiffness measurement from sinusoidal mode beam bending experiment**

Fig S6 shows the contact stiffness evolution during beam bending experiment performed on PLD a-$Al_2O_3$ films. Continuous Stiffness Mode (i.e. sinusoidal oscillation) was superimposed on the loading segment during the test. The contact stiffness increases initially due to indenter penetration into the beam (zone I). Thereafter, it decreases due to sliding of the tip on the top of the beam away from the fixed end (zone II). Finally, the contact stiffness increases further beyond 2 μm due to rapid increase in indentation depth and decrease in sliding (zone III).

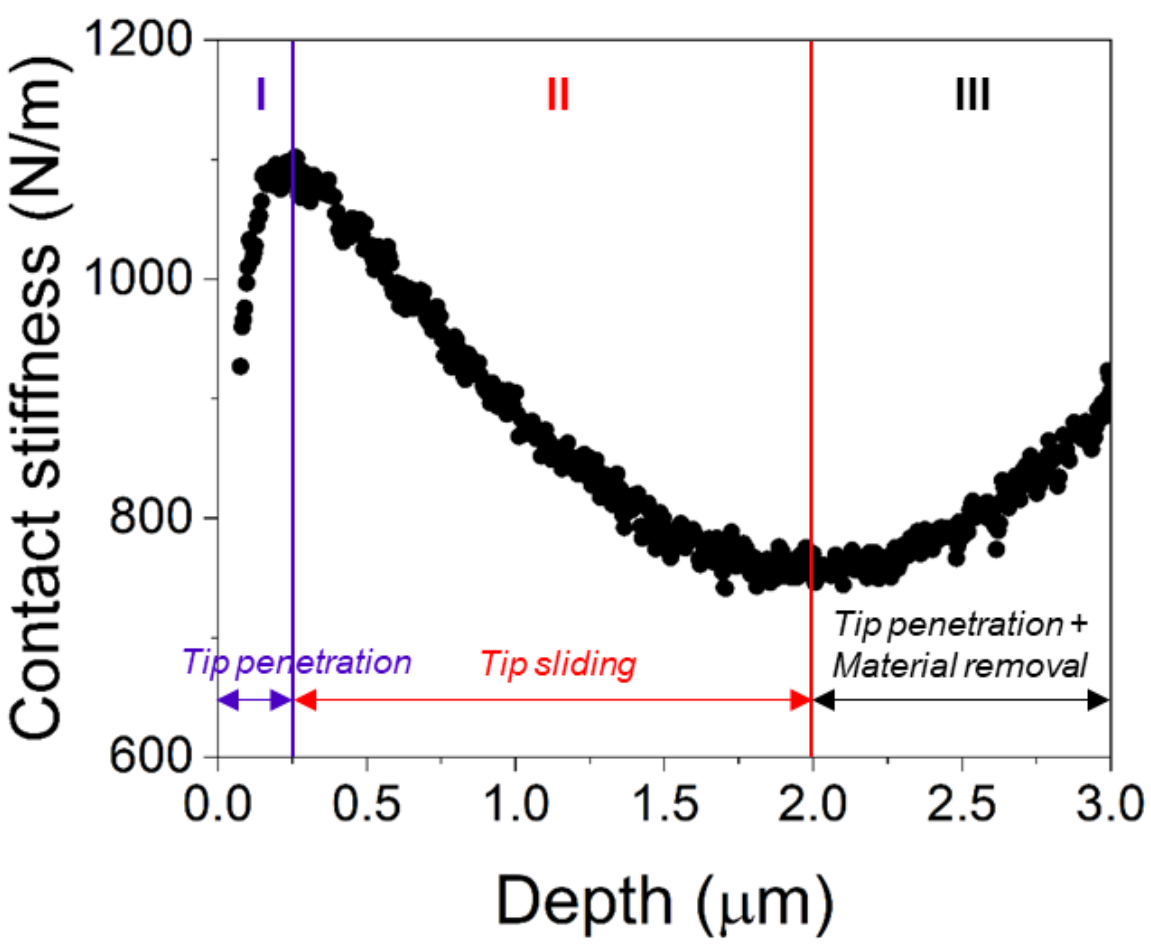

*Fig. S6: Contact stiffness evolution as a function of indenter displacement during beam bending. The contact stiffness is calculated from a small sinusoidal signal that was imposed during the bending test.*

**S5. Microcantilever fracture tests of a-$Al_2O_3$ films**

The raw load-displacement curves of the microcantilever fracture experiments of all a-$Al_2O_3$ films are shown in Fig. S7. All tests were performed at a constant tip velocity of 10 nm/s maintaining a similar loading distance for the cantilevers. The films showed linear elastic response before the brittle failure of films.

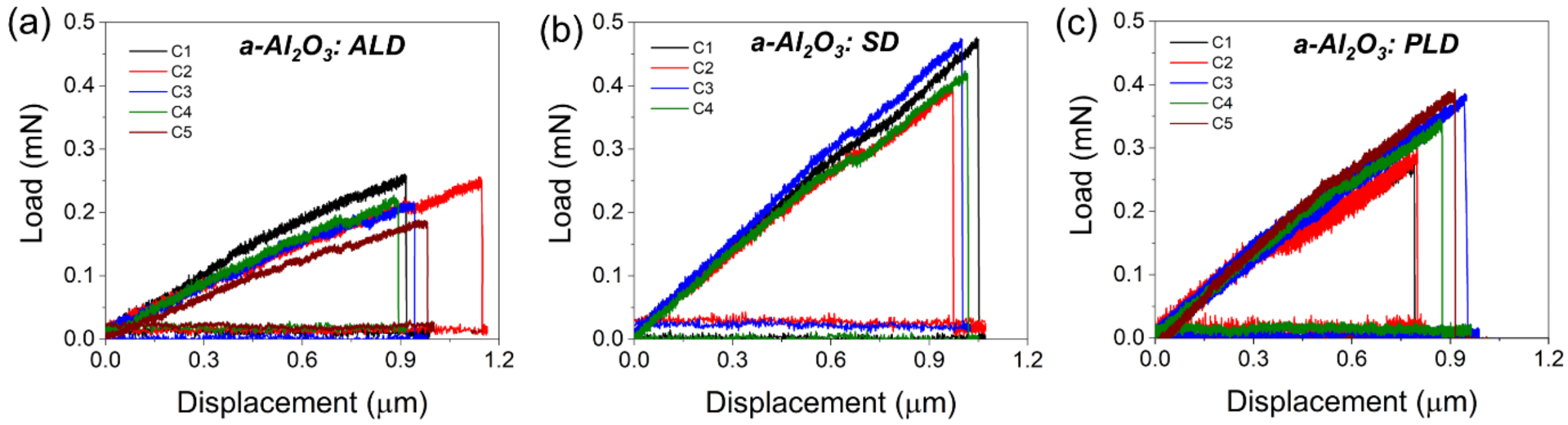

*Fig. S7: Representative load-displacement curves of a) ALD, b) SD and c) PLD a-$Al_2O_3$ films during notched microcantilever fracture experiments.*

**S6. Cross-section analysis of alumina films for defect analysis**

Cross-section analysis of alumina films in SEM did not reveal any microscale defects in the films as all of them appeared completely dense (Fig. S8). However, detailed investigation of SD films in TEM showed nanometer sized linear features (Fig. S9). These same features are not visible in the corresponding dark-field image, which rules out the possibility of crystalline inclusions or oriented nanocrystalline phases that would otherwise produce diffraction contrast. The linear morphology and alignment of these features along the film growth direction further supports their identification as inter-columnar voids formed during SD deposition, a characteristic signature of the columnar microstructure evolution

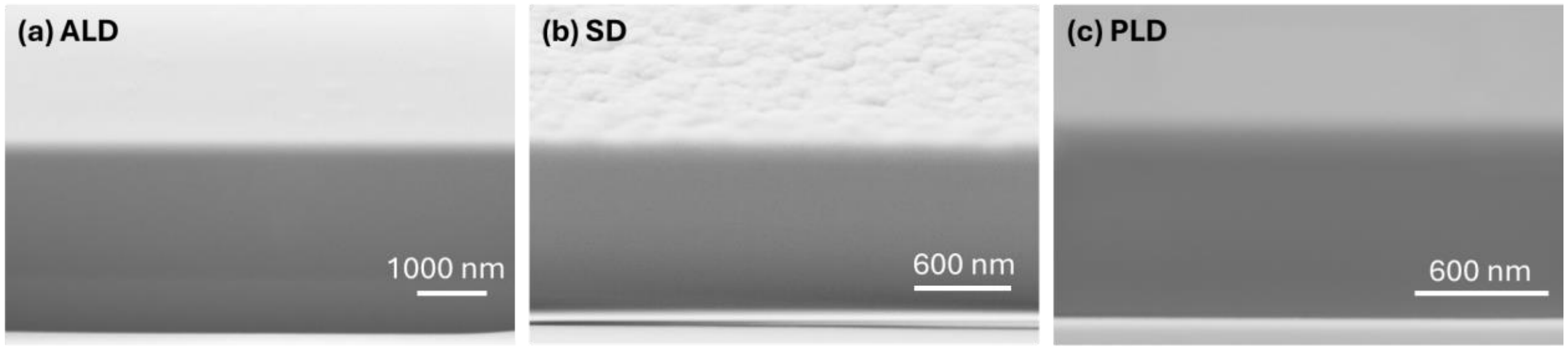

*Fig. S8. TEM images showing the cross-section of a) ALD, b) SD and c) PLD alumina films*

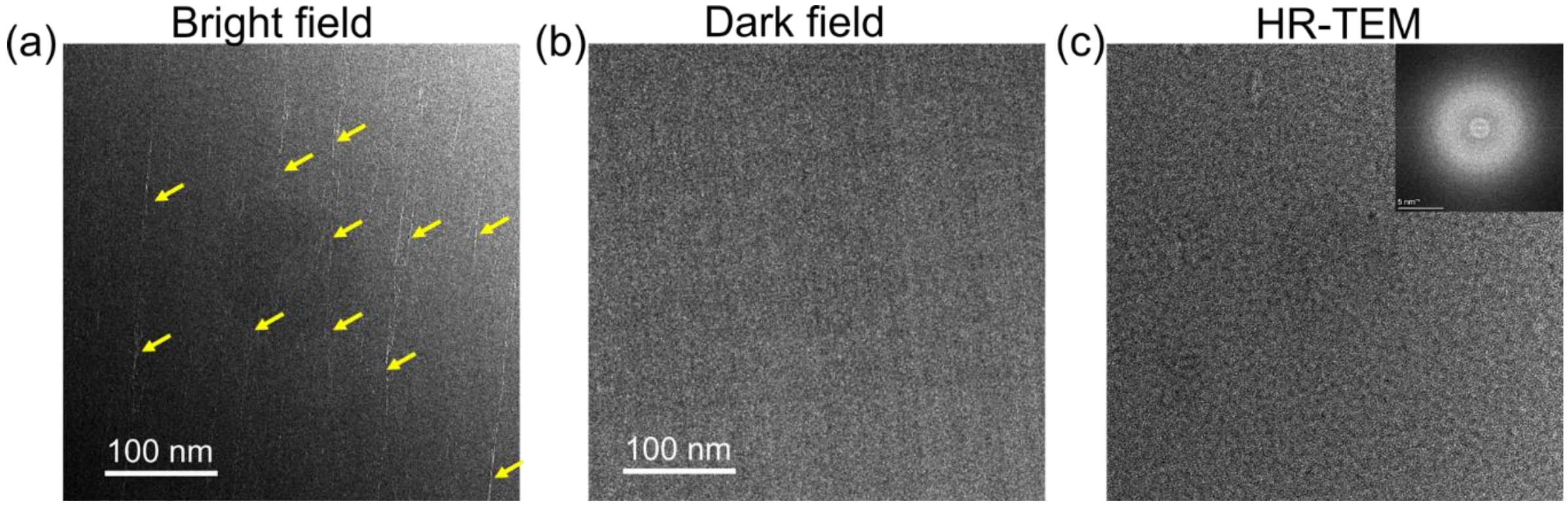

*Fig. S9. TEM images showing the a) bright filed, b) dark field and c) HR-TEM images of SD alumina sample. (Columnar voids are marked in yellow arrows in (a)).*

**S7. Effect of TEM (200 kV) electron beam on the PLD alumina**

The TEM lamella of PLD micropillar sample was observed under the TEM to identify whether any structural change had occurred during the deformation. A fully amorphous structure, as seen in Fig S10a, was observed where no crystallization had occurred due to $Ga^{+}$ ion influence. However, the formation of crystallites was favored in the regions of lamella under e-beam exposure for more than 60s. The TEM diffraction pattern, as seen in Fig S10b, shows the bright spots indicating the presence of a crystalline phase.

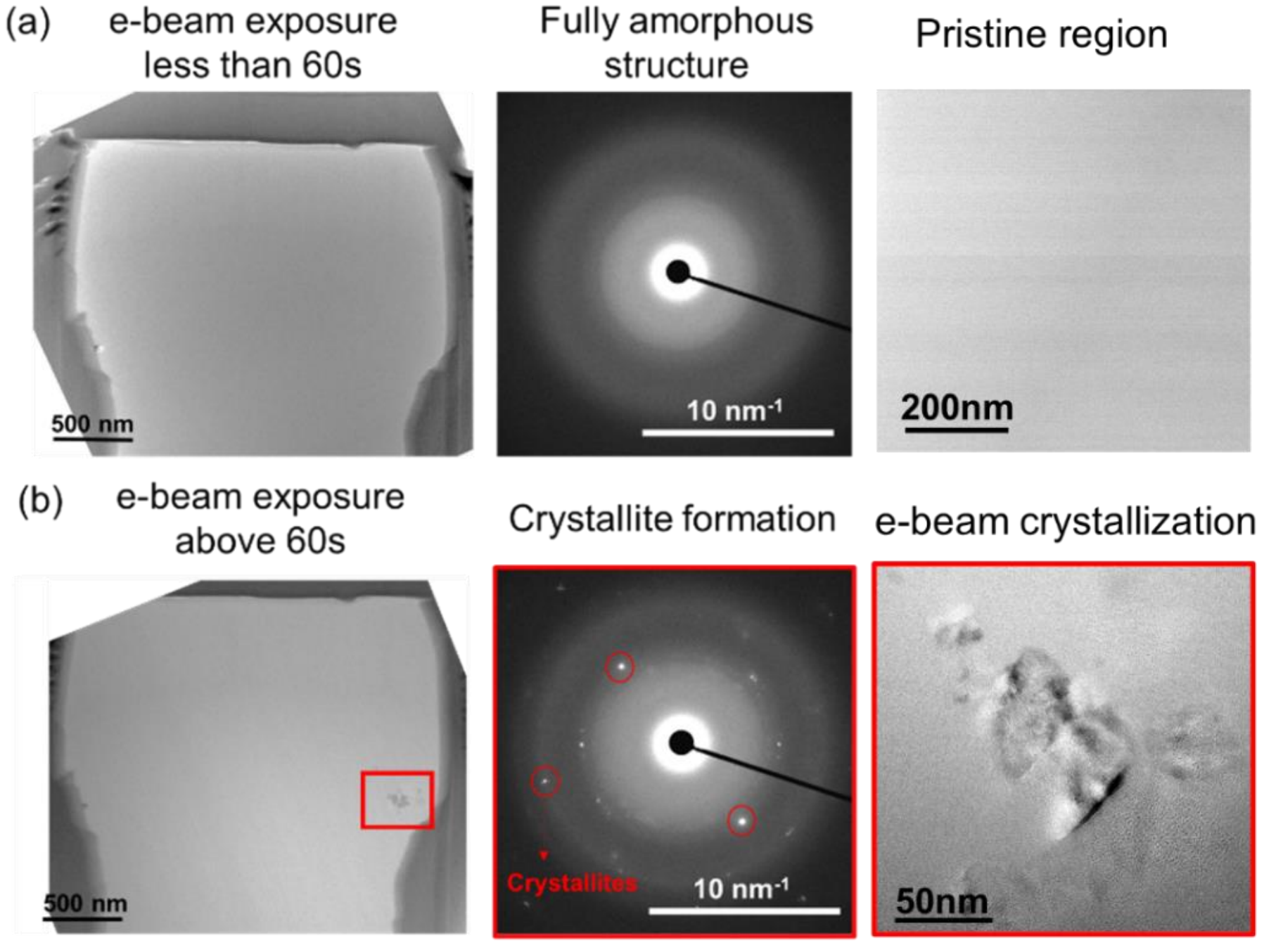

*Fig. S10: TEM image and corresponding diffraction of PLD alumina micropillar showing a) fully amorphous structure, and b) crystallite formation as a result of e-beam exposure*

**S8. Effect of SEM (10 kV) electron beam on the PLD alumina**

The influence of SEM electron beam in altering the mechanical behavior of amorphous $Al_2O_3$ film during the *in situ* SEM micromechanical testing was studied. Micropillar compression experiments on ALD, PLD and SD amorphous alumina were conducted under e-beam ON and OFF conditions. It was observed that the engineering stress-strain response of the micropillars were not influenced by the e-beam during testing, as similar yield strength value was observed for each material in both ON and OFF conditions as seen in Fig S11.

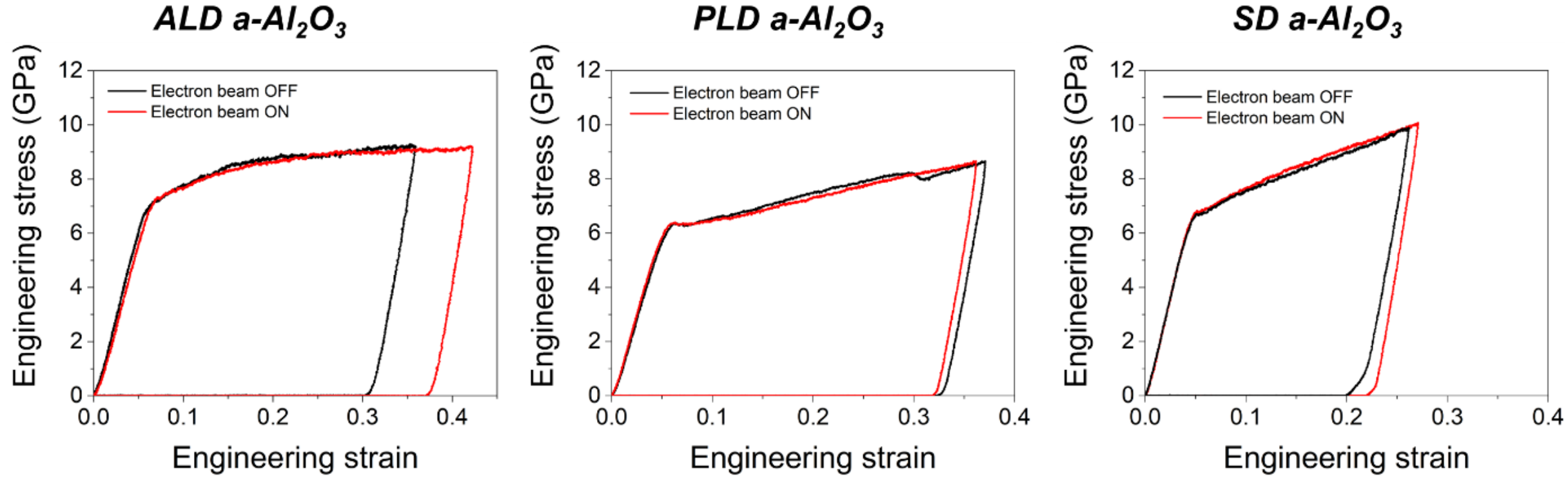

*Fig. S11: Micropillar compression results on ALD, PLD and SD amorphous alumina micropillar samples under e-beam ON and OFF conditions*